\documentclass[pre,
  twocolumn,
  twoside,
  byrevtex,
  superscriptaddress,
  floatfix,
  longbibliography,
]{revtex4-1}

\usepackage{kotex}

\usepackage[section]{placeins}

\usepackage{currfile}

\PassOptionsToPackage{hyphens}{url}\usepackage{hyperref}
\makeatletter
\g@addto@macro{\UrlBreaks}{\UrlOrds}
\makeatother

\hypersetup{
  colorlinks=true,
  allcolors=todoblue,
  urlcolor=todoblue,
  citecolor=todoblue,
  pdfborder={0 0 0},
  breaklinks=true,
}

\newcommand{\firsttwitterdate}{2008/09/09}

\DeclareRobustCommand{\firstpolldate}{1941/07/22\unskip}
\DeclareRobustCommand{\lastpolldate}{2020/12/17\unskip}

\DeclareRobustCommand{\lasttwitterdate}{2021/10/24\unskip}
\DeclareRobustCommand{\lasttwitteryear}{2021\unskip}
\DeclareRobustCommand{\lowtrumprank}{1541\unskip}
\DeclareRobustCommand{\lowtrumpdate}{2021/04/21\unskip}
\DeclareRobustCommand{\trumpmedianrank}{191\unskip}
\DeclareRobustCommand{\godmedian}{303\unskip}
\DeclareRobustCommand{\godcione}{237\unskip}
\DeclareRobustCommand{\godcitwo}{383\unskip}
\DeclareRobustCommand{\godtwofive}{281\unskip}
\DeclareRobustCommand{\godsevenfive}{330\unskip}
\DeclareRobustCommand{\godmin}{134\unskip}
\DeclareRobustCommand{\godmax}{529\unskip}

\DeclareRobustCommand{\oufobama}{2.5\unskip\%}
\DeclareRobustCommand{\oufmccain}{1.0\unskip\%}
\DeclareRobustCommand{\oufromney}{0.1\unskip\%}
\DeclareRobustCommand{\oufhillary}{0.8\unskip\%}
\DeclareRobustCommand{\ouftrump}{33.9\unskip\%}
\DeclareRobustCommand{\oufbtstwt}{50.0\unskip\%}

\usepackage{color}

\definecolor{todoblue}{RGB}{0, 91, 187}
\newcommand{\todo}[1]{\noindent\textcolor{todoblue}{{$\Box$ #1}}}

\definecolor{olivegreen}{rgb}{0.33333,.41961,0.18431}
\definecolor{forestgreen}{rgb}{0.13333,.5451,0.13333}
\definecolor{lightgrey}{rgb}{0.7,0.7,0.7}
\definecolor{verylightgrey}{rgb}{0.90,0.90,0.90}
\definecolor{grey}{rgb}{0.5,0.5,0.5}
\definecolor{gray}{rgb}{0.5,0.5,0.5}

\usepackage{graphicx,epsfig,verbatim,enumerate}
\usepackage{amssymb,amsmath}
\usepackage{ifthen}

\usepackage{longtable}

\usepackage{array}

\usepackage{mathtools}

\newcommand{\med}{\operatorname{\textnormal{med}}}

\usepackage{textgreek}

\newboolean{twocolswitch}

\newcommand{\sindex}[1]{}
\newcommand{\nindex}[1]{}

\newcommand{\www}[1]{\url{#1}}

\usepackage{lettrine}

\newcommand{\termcount}{N}
\newcommand{\termrate}{R}
\newcommand{\termrelrate}{\termrate^{\textnormal{rel}}}

\newcommand{\termratemax}{\termrate^{\textnormal{max}}}

\newcommand{\termcountfn}[2]{\termcount_{#1,#2}}
\newcommand{\termratefn}[2]{\termrate_{#1,#2}}
\newcommand{\termrelratefn}[2]{\termrelrate_{#1,#2}}

\newcommand{\zipfrank}{r}

\setboolean{twocolswitch}{true}

\begin{document}

\title{\protect
Fame and Ultrafame:
Measuring and comparing daily levels of `being talked about' for
United States' presidents,
their rivals,
God,
countries,
and K-pop
}

\author{
  \firstname{Peter Sheridan}
  \surname{Dodds}
}

\email{peter.dodds@uvm.edu}

\affiliation{
  Computational Story Lab,
  Vermont Complex Systems Center,
  MassMutual Center of Excellence for Complex Systems and Data Science,
  Vermont Advanced Computing Core,
  University of Vermont,
  Burlington, VT 05401.
  }

\affiliation{
  Department of Mathematics \& Statistics,
  University of Vermont,
  Burlington, VT 05401.
  }

\author{
  \firstname{Joshua R.}
  \surname{Minot}
}

\affiliation{
  Computational Story Lab,
  Vermont Complex Systems Center,
  MassMutual Center of Excellence for Complex Systems and Data Science,
  Vermont Advanced Computing Core,
  University of Vermont,
  Burlington, VT 05401.
  }

\author{
  \firstname{Michael V.}
  \surname{Arnold}
}

\affiliation{
  Computational Story Lab,
  Vermont Complex Systems Center,
  MassMutual Center of Excellence for Complex Systems and Data Science,
  Vermont Advanced Computing Core,
  University of Vermont,
  Burlington, VT 05401.
  }

\author{
  \firstname{Thayer}
  \surname{Alshaabi}
}

\affiliation{
  Computational Story Lab,
  Vermont Complex Systems Center,
  MassMutual Center of Excellence for Complex Systems and Data Science,
  Vermont Advanced Computing Core,
  University of Vermont,
  Burlington, VT 05401.
  }

\author{
  \firstname{Jane Lydia}
  \surname{Adams}
}

\affiliation{
  Computational Story Lab,
  Vermont Complex Systems Center,
  MassMutual Center of Excellence for Complex Systems and Data Science,
  Vermont Advanced Computing Core,
  University of Vermont,
  Burlington, VT 05401.
  }

\author{
  \firstname{David Rushing}
  \surname{Dewhurst}
}

\affiliation{
  Computational Story Lab,
  Vermont Complex Systems Center,
  MassMutual Center of Excellence for Complex Systems and Data Science,
  Vermont Advanced Computing Core,
  University of Vermont,
  Burlington, VT 05401.
  }

\author{
  \firstname{Andrew J.}
  \surname{Reagan}
}

\affiliation{
  MassMutual Data Science,
  Amherst,
  MA 01002.
  }

\author{
  \firstname{Christopher M.}
  \surname{Danforth}
}

\affiliation{
  Computational Story Lab,
  Vermont Complex Systems Center,
  MassMutual Center of Excellence for Complex Systems and Data Science,
  Vermont Advanced Computing Core,
  University of Vermont,
  Burlington, VT 05401.
  }

\affiliation{
  Department of Mathematics \& Statistics,
  University of Vermont,
  Burlington, VT 05401.
  }

\date{\today}

\begin{abstract}
  \protect
  When building a global brand of any kind---a political actor, clothing style,
or belief system---developing widespread awareness is a primary goal.
Short of knowing any of the stories or products of a brand,
being talked about in whatever fashion---raw fame---is,
as Oscar Wilde would have it,
better than not being talked about at all.
Here, we measure, examine, and contrast the day-to-day
raw fame dynamics on Twitter
for US Presidents and major US Presidential candidates from 2008 to 2020:
Barack Obama,
John McCain,
Mitt Romney,
Hillary Clinton,
Donald Trump,
and
Joe Biden.
We assign ``lexical fame''
to be the number and (Zipfian) rank of the (lowercased) mentions made for each
individual across all languages.
We show that all five political
figures have at some point reached extraordinary volume levels
of what we define to be 
``lexical ultrafame'':
An overall rank of approximately 300
or less which is largely the realm of function words and demarcated by
the highly stable rank of `god'.
By this measure, `trump'
has become enduringly ultrafamous, from the 2016 election on.
We use typical ranks for country names and function words as standards
to improve perception of scale.
We quantify relative fame rates and find that in the eight
weeks leading up the 2008 and 2012 elections, `obama' held a 1000:757
volume ratio over `mccain' and  1000:892
over `romney',
well short of the
1000:544
and
1000:504
volumes favoring `trump'
over `hillary'
and `biden'
in the 8 weeks leading up to the
2016 and 2020 elections.
Finally, we track how only one other entity has more sustained
ultrafame than `trump' on Twitter: The K-pop (Korean pop) band BTS.
We chart the dramatic rise of BTS, finding
their Twitter handle `@bts\_twt' has been able to compete
with `a' and `the', reaching a rank of three at
the day scale and a rank of one at the quarter-hour scale.
Our findings for BTS more generally point to K-pop's growing
economic, social, and political power.


\end{abstract}


\maketitle

\section{Introduction}
\label{sec:ultrafame.introduction}

\begingroup
\advance\leftmargini -7pt
\begin{quotation}
\noindent
\textnormal{``It is silly of you, for there is only one thing in the world worse
than being talked about, and that is not being talked about.''}

{\small --- Oscar Wilde, The Picture of Dorian Gray~\cite{wilde1890a}}.
\end{quotation}
\endgroup

``Being talked about'' is the essence of fame,
a word that accurately encodes this most basic of sociological mechanisms
as it traces back to
the Latin f\={a}ma (``speak'')
with 
\textphi{$\acute{\mbox{\texteta}}$}\textmugreek\texteta\ 
(ph\'{\={e}}m\={e}, ``talk'') as its Greek cognate.

Achieving widespread awareness is arguably the primary goal
of any people-centric enterprise seeking to scale.
Of course any such enterprise will want the valence of fame
to be positive, and for ``talk'' to be self-sustaining.
Examples abound.
To take just one, in the sphere of sport,
Lance Armstrong's archetypal fall-from-grace
followed a global expansion of awareness of
cancer research, the Tour de France, and cycling.
Armstrong himself became famous as an eight-fold kill-the-monster hero,
first conquering cancer then the Tour seven times in a row,
all ending with a televised confession of betrayal to Oprah.

We also know that fame is profoundly a social construct,
a complex mix of system randomness, an individual's luck, timing, history, 
and, to the extent that it exists at all in a given field,
inherent quality~\cite{salganik2006a,salganik2008a,dodds2013b}.
From the perspective of collective evaluation of cultural entities,
the existence and perceived importance
of ranked lists of anything (wealthy individuals, songs, books, colleges, cities, countries)
leaves social systems
vulnerable to those unethical actors who would seek fame.
Knowing that ``getting the word out there'' is the foundational work
allows system-level manipulation by individuals or organizations pretending
to be at or near the top of such lists
by gaming myriad sociotechnical algorithms
(many/some ``people are saying''~\cite{cookson2018a,wikipedia-weaselwords2019a},
payola~\cite{coase1979a},
``John Barron''~\cite{greenberg2018a,borchers2016a,wikipedia-pseudonyms-of-donald-trump2019a}).

\begin{figure}[!tp]
  \includegraphics[width=\columnwidth]{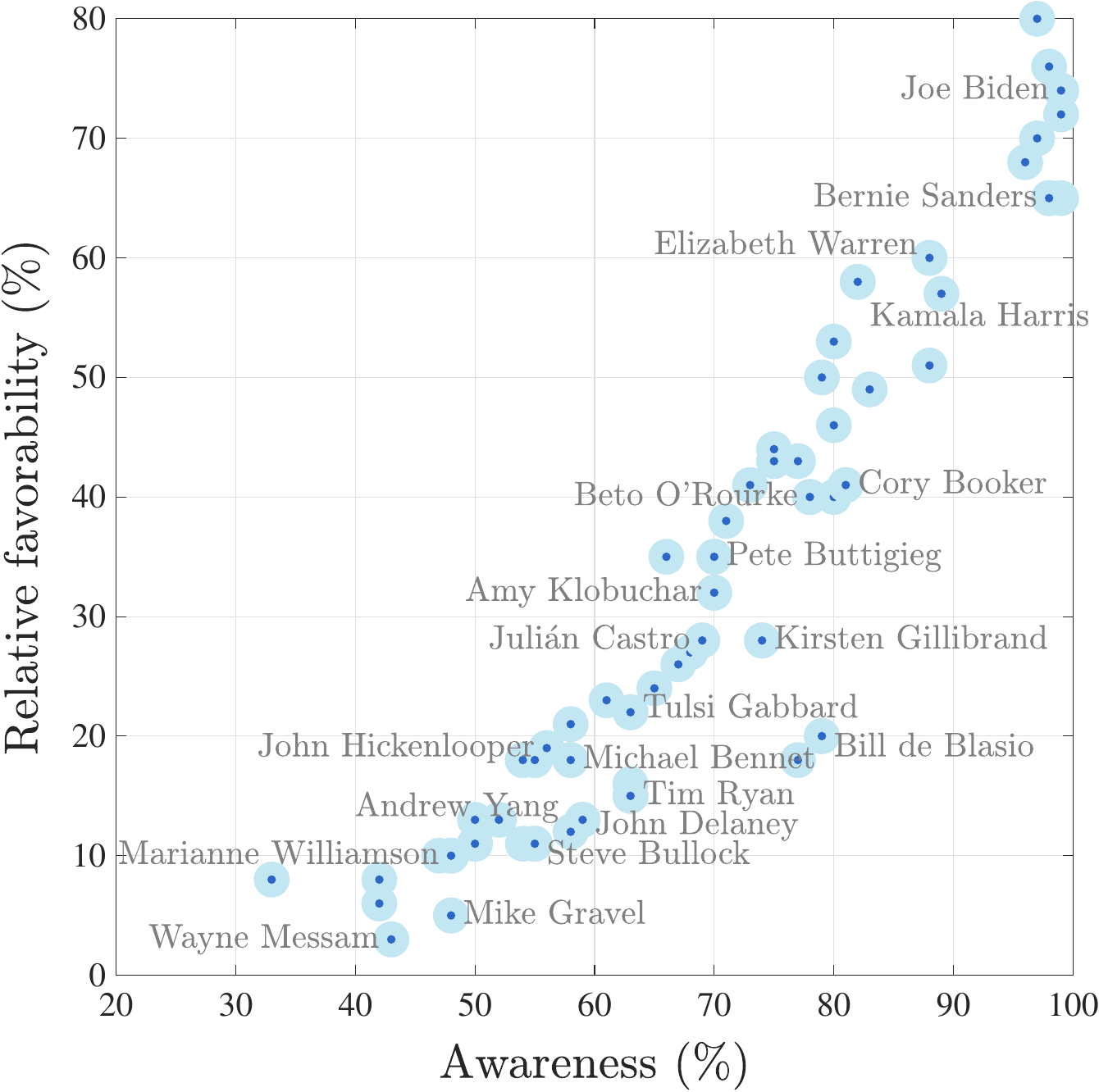}
  \caption{
    Comparison of awareness and relative favorability for 24 democratic candidates for
    the democratic nominee in the 2020 US presidential election, showing a strong
    positive correlation (Spearman correlation coefficient $r_{\textnormal{s}}$=0.949).
    Details:
    The data comes from four polls carried out by Monmouth University
    from 2019/01 to 2019/05~\cite{monmouthpoll2019-02-04a}.
    We compute a candidate's
    relative favorability as normalized by the subset of respondents who have heard of that candidate.
    Not all candidates were included in all polls resulting in 58 data points (instead of 96).
    We acknowledge that the varying numbers of polls per candidate calls for a more sophisticated
    analysis than simple correlation, but our aim is simply to show a clear
    example of well correlated awareness and favorability.
    See Fig.~\ref{fig:ultrafame.president_favorability002} for a counterexample.
    For readability, we only show a subset of unique names,
    and arrange these left and right so that one end of the text is close to the relevant data point.
  }
  \label{fig:ultrafame.monmouth002}
\end{figure}

In politics, a key polling question concerns whether
or not an interviewee has heard of a candidate \textit{at all}---shorn of sentiment and story.
While some polls show that increases in awareness correspond to increases
in favorability,
politicians trace out many paths in awareness-favorability space.

For example,
as we show in Fig.~\ref{fig:ultrafame.monmouth002},
a series of polls carried out by Monmouth University during
the first five months of 2019~\cite{monmouthpoll2019-02-04a}
revealed a strong correlation between
awareness of
and
favorability
toward 24 potential Democratic candidates for the 2020 presidential
election
(Spearman correlation coefficient: $r_{\textnormal{s}}$=0.949).
The awareness extremes were for Joe Biden,
who registered 1\% of those polled saying
they had not heard of him (2019/05), and 67\% saying the same
of Marianne Williamson (2019/03).

\begin{figure}[!tp]
  \includegraphics[width=\columnwidth]{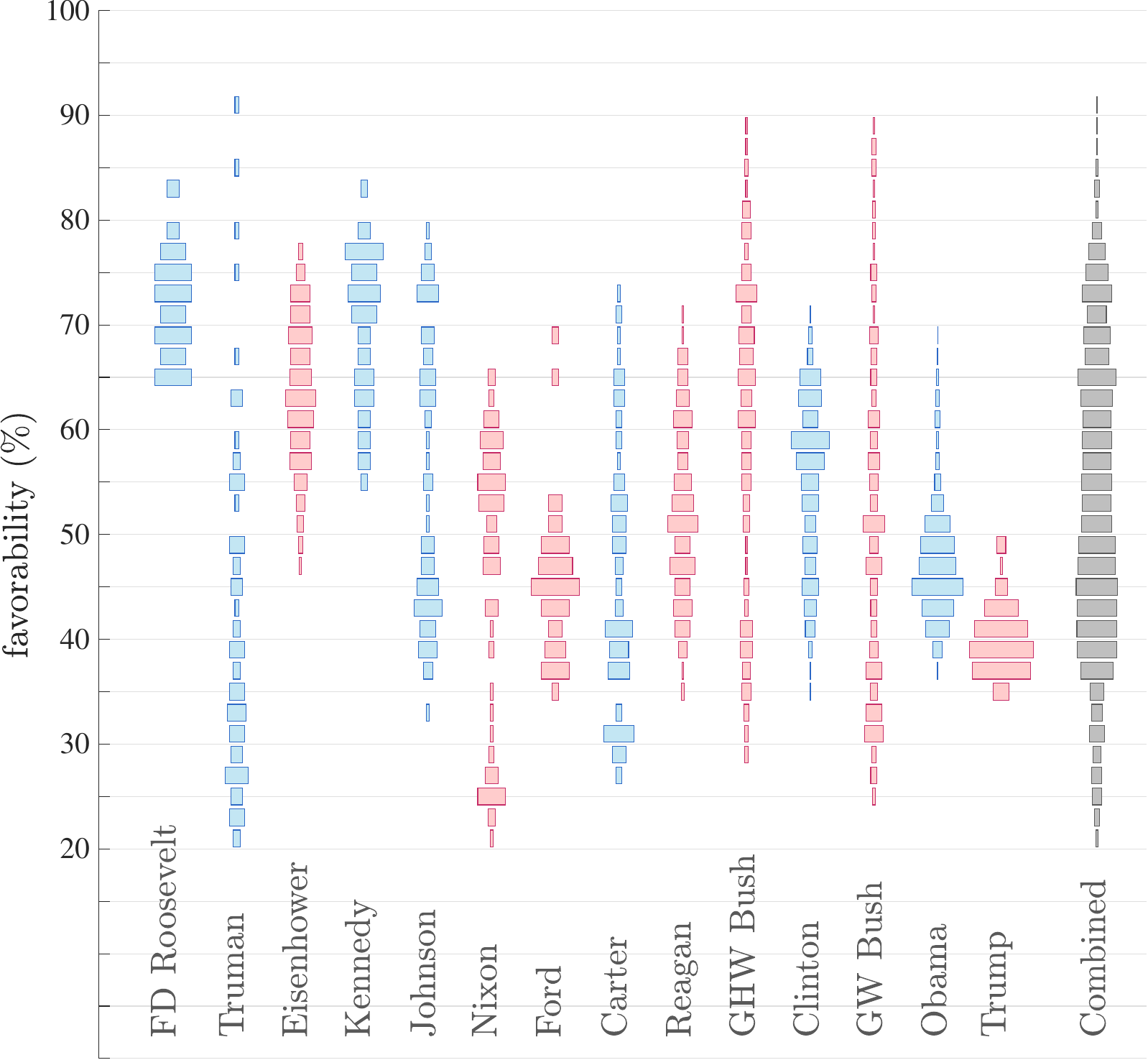}
  \caption{
            Histograms of favorability ratings from Gallup polls
    for US presidents taken from Franklin Roosevelt through to Donald Trump~\cite{woolley2008a}.
    The rightmost histogram represents a combined average with each president's ratings
    equally weighted.
    We do not have data for awareness levels but for US presidents we can
    reasonably assert that the percentage will
    be uniformly high.
    In contrast to the strong correlation between awareness and favorability in Fig.~\ref{fig:ultrafame.monmouth002},
    we see that high-awareness political figures can certainly achieve a wide
    range of favorability ratings.
    With our focus in this paper on awareness---raw fame---we offer this figure as
    a tempering exhibit.
    Polls run from \firstpolldate\ through to \lastpolldate.
    Trump has had the lowest variation in favorability
    (variance $\sigma^{2}$=3.05, 2.5 to 97.5 percentile covering 14.0 favorability points)
    while Truman has the highest 
    (variance $\sigma^{2}$=16.0,  2.5 to 97.5 percentile covering 63.1 favorability points).
    Trump is also the only president to have not registered an above 50 favorability rating
    in Gallup polls.
  }
  \label{fig:ultrafame.president_favorability002}
\end{figure}

By contrast,
as we show in Fig.~\ref{fig:ultrafame.president_favorability002},
US presidents provide a powerful example as figures with
extremely high global awareness levels
while receiving a wide variation of approval over time and across demographics~\cite{woolley2008a}.
Nevertheless and as for many other cultural and social spheres,
achieving widespread ``brand awareness''
in politics is an order zero undertaking.

Our intention with Figs.~\ref{fig:ultrafame.monmouth002}
and~\ref{fig:ultrafame.president_favorability002}
is to show that while simple awareness can indeed correlate with favorability as a brand grows, this is not a general truth. 

So, while exploring mechanisms, sentiment, narratives,
and other aspects of fame are all
necessary~\cite{rosen1981a,adler1985a,balinski2007a,laureti2004b,chase2002a,salganik2006a,salganik2008a,dodds2013b},
we will here concern ourselves with Wildean raw fame---the state of being talked about---for
US presidents and their main rivals.

\begin{table*}[tp!]
  \bgroup
  \def\arraystretch{1.75}%
  \begin{tabular}{
      |l|l|l|
    }
    \hline
    Political~figure:
    &
    Position:
    &
    1-grams (dominant 1-gram in \textbf{bold}):
    \\
    \hline
    \hline
    Barack Obama
    &
    US~president from 2009/01 to 2017/01
    &
    barack, \textbf{obama}, @barackobama
    \\
    \hline
    John McCain
    &
    Republican Party nominee in 2008
    &
    john, \textbf{mccain}, @sejohnmccain
    \\
    \hline
    Mitt Romney
    &
    Republican Party nominee in 2012
    &
    mitt, \textbf{romney}, @mittromney
    \\
    \hline
    Hillary Clinton
    &
    Democratic Party nominee in 2016
    &
    \textbf{hillary}, clinton, @hillaryclinton
    \\
    \hline
    Donald Trump
    &
        US~president from 2017/01 to present
    &
    donald, \textbf{trump}, @realdonaldtrump
    \\
    \hline
    Joe Biden
    &
    Democratic Party nominee in 2020
    &
    joe, \textbf{biden}, @joebiden
    \\
    \hline
  \end{tabular}
  \egroup
  \caption{
    The six political figures whose
    fame we trace and compare via Twitter mentions.
    To quantify fame,
    we measure the rank and count dynamics for three 1-grams
    for each political figure:  
    First name, last name, and Twitter handle.
    The 1-grams in bold are the on-average, unambiguous 1-gram with the highest count
    referring to the political figure.
    We also follow the lexical fame of the
    K-pop band BTS~\cite{BTSname} per their Twitter handle
    @bts\_twt.
  }
  \label{tab:ultrafame.entities}
\end{table*}

We focus on the major political figures involved
in the US presidential
elections of 2008, 2012, 2016, and 2020,
and the encompassing time frame:
Barack Obama,
John McCain,
Mitt Romney,
Hillary Clinton,
Donald Trump,
and
Joe Biden
(see Tab.~\ref{tab:ultrafame.entities}).

As we will show for these political figures,
Trump has enjoyed singularly transcendent fame---what we will call lexical ultrafame---within
our time period of interest.
Trump has only been overmatched by one other entity of any kind:
The K-pop (Korean pop) band BTS~\cite{BTSname},
and we find that we are obliged to include them in our analyses.
BTS's ultrafame more generally reflects the real power that K-pop and K-pop fandom
has attained as a global cultural, social,
and political force~\cite{aisyah2017a,dal2018a,anderson2018a,wikipedia-bts-impact2019a}.
Extraordinarily, both Trump and BTS have occupied levels of lexical fame
that the most basic function words of language occupy.
To give a sense of what we will uncover,
the median rank (which we explain below)
for the word `trump'
during Trump's presidency has been around $\zipfrank$=180, akin to
that of words like
`after',
`would',
and
`man',
while the median rank for BTS's Twitter handle, `@bts\_twt',
has tracked near $\zipfrank$=90, a rank typical of
the words
`has',
`more',
and `da'.

There are many ways to gauge fame
such as 
direct polls,
mentions on social media,
and
rates of internet searches.
While including an array of distinct measures would be ideal,
we limit ourselves here to the social media platform that is Twitter.
We will thus endeavor to perform our analyses with great care
for one well-defined, if sprawling,
realm of public discourse.

To be explicit, Twitter is of course just one large 
online space 
that provides a single source of text data for measuring the relative prevalence of terms. 
While we would argue that Twitter is notable for its reach and observed impact on social and political systems, we would never suggest that measurements of fame derived from the platform are all encompassing.
As will show, our measurement of day-scale fame time series on Twitter functions well for generating computationally-aided historical timelines.
It is our hope that our in-depth analysis of fame on Twitter will serve as a touch point for other measurements of fame (e.g., using Google Trends, Wikipedia, books, etc.).

For our purposes here, we will define lexical fame
of any given entity by the daily counts and
Zipfian ranks for 1-grams (words, hashtags, user handles, etc.)
pertaining to that entity.
For example, Barack Obama's lexical fame will be registered
by counts and ranks for `barack', `obama', and `@barackobama'.
For a few 1-grams on specific days,
we will report on fame levels at the 15 minute time scale.
We also limit ourselves to 1-grams, reserving full analyses of
$n$-grams for $n \ge 2$ for future work, though we will mention
a few observations for 2-grams for specific days.

We deliver the remainder of our paper as follows.
In Sec.~\ref{sec:ultrafame.data}
we describe our Twitter data set
and the data-wrangling part of our analysis,
reserving details for Sec.~\ref{sec:ultrafame.methods} at the end.
We present our core results in Sec.~\ref{sec:ultrafame.results}.
In Sec.~\ref{subsec:ultrafame.timeseries},
we first examine time series and histograms for ranks of Twitter mentions
for our six political figures and the K-pop band BTS.
In Sec.~\ref{subsec:ultrafame.comparisons},
we then make comparative analyses of mentions across figures and
across calendar years and the eight weeks leading up to
the US\ elections.
Our work is observational and descriptive---a fundamental aspect of basic science---and we will not
move toward prediction here.
We close with concluding remarks in
Sec.~\ref{sec:concludingremarks}.
We have also constructed our figures and captions to be
as self-contained as possible for those readers who
may not wish to read the main text,
and by inclusion, this sentence.

\section{Data and preparatory treatment}
\label{sec:ultrafame.data}

\subsection{Description of Twitter data set and rationale for use}
\label{subsec:ultrafame.twitter}

We measure the daily fame of political figures as reflected by
mentions on the
social media platform Twitter.
We have collected roughly 10\% of all public tweets starting on \firsttwitterdate\ through to \lasttwitterdate,
allowing us to explore
fame dynamics around the last three US presidential elections.

Twitter has a number of well known benefits and drawbacks.
First, a few of the stronger positives.
We have essentially real-time temporal resolution for a massive scale
of messages.
Standardization of hashtags 
have made for powerful codifying of issues (e.g., \#metoo),
and formalization of retweets, favorites, and replies allow us to follow
the reaction to individual tweets in detail.
Though accounts can be made private,
Twitter is by default public-facing and
largely engaged with such an understanding by its users.
The world's languages, and not just the dominant ones,
are all present on Twitter, allowing for potentially rich
cultural and linguistic explorations.

Negatives for Twitter are also on offer in good number.
Twitter is not used uniformly by all people around the world,
with users skewing younger and democratic, 
and engaging at a wide range of rates, 
and with strong user bases in,
for example, the US, Japan, and Brazil~\cite{wojcik201a}.
Geolocation and demographic features have typically been
publically available for a
small fraction of tweets (less than 1\% for the former).
Geolocation has been uneven in nature (latitude-longitude versus place name),
and was removed entirely as a feature for users in 2019
though metadata in photos could still encode location.
Algorithmically generated content is prevalent (e.g., ``bots'')
and problematic for the both the
service and users~\cite{ferrara2016a}.
The changing nature of how Twitter presents information to user
through algorithmic feeds and trending story pages only adds
further complexity.

In the middle lies the evident issue that tweets,
and cleverly constructed subsets of tweets,
do not perfectly represent all the ideas, viewpoints, and utterances of
people of whatever category one may want to study.
The collective voice of Twitter is a
discordant symphony
of the expressions, reactions, and amplifications
of
individuals,
news outlets,
corporations,
fan bases,
celebrities,
and
automatic accounts of all alignments.
The amplification processes are
rich-get-richer mechanisms~\cite{simon1955a,price1976a}
made possible by follower networks, external media's embedding
of tweets, and Twitter's own system of curating and presenting
trending stories.

We know that as a whole Twitter strongly follows major events~\cite{dodds2011e,becker2011a,dodds2020n}
and can successfully be used as an indirect polling system~\cite{dodds2011e,cody2016b,alshaabi2021b,alshaabi2021c}.
Twitter has also risen in prominence in the political sphere, particularly with the usage
of the platform by the former US president, Donald Trump.
In turbulent times, recalling what major events occurred and in what order temporally
can be challenging---``chronopathy''---and we have been able to use Twitter
to generate computational timelines around Trump~\cite{dodds2020n}.
With daily (and sub-day) resolution of lexical fame,
we find here that we are able, by inspection, to tie rank dynamics to specific events.

Like other global social media giants of today, 
Twitter has the potential to create real impact at all scales.
Of many examples, one thematically related
to our study here is the identification of President Trump's tweets
as having an effect on prices of Treasury bonds,
leading JP Morgan Chase to create a
covfefe-fueled
``Volfefe index''~\cite{wikipedia-volfefe-index2019a}
(see also~\cite{clarke2019a}).
Entwining news, politics, markets, patriotism issues, and belief,
a 2013 hacked tweet from the Associated Press's Twitter account
suggesting that the White House had been bombed and Obama was injured,
leading to an immediate drop in the market~\cite{washingtonpost2013-04-23ap-tweet-obama}.
Although the story was quickly corrected, this one hacked tweet
caused the evaporation of \$136 billion in a few minutes.
One more example, this time showing the power of a celebrity's off-handed remark:
On February 21, Kylie Jenner, tweeted, and we quote,
``sooo does anyone else not open Snapchat anymore? Or is it just me\ldots\ ugh this is so sad''~\cite{guardian2018-02-23snapchat}.
Subsequent to this single tweet---though we do not here claim causality---Snapchat's shares deflated 6\% in value (\$1.3 billion).

We are not implying that 
Twitter is solely dominated by individual tweets.
Words, phrases,  and hashtags that reflect major events and stories going on in the world
will be produced and amplified collectively
(e.g., `coronavirus', \#blacklivesmatter)~\cite{alshaabi2021c}.

In short, Twitter is a large-scale,
temporally fine-grained
source of written text
containing meaningful signatures
that can powerfully affect society and the world.
Even shorter, 
Twitter is Twitter.

\subsection{Preparation of Twitter data set for analysis}
\label{subsec:ultrafame.analysis}

To explore raw fame and ultrafame,
we take our entire Twitter corpus
and process
tweets into 1-grams.
While keeping the parsing as simple as possible, we make
some choices
such as discarding emojis,
excluding languages that do not use whitespace,
and adjusting all letters to lower case for languages where two cases exist
(e.g., counts for `god' include counts for `God', `GOD', `god', etc.)
(see Methods for full details, Sec.~\ref{sec:ultrafame.methods}).
Such parsing is evidently not an activity that humans could perform, and even if they could,
reading (or perhaps more accurately, absorbing) a stream of
50 million tweets a day could well be harmful
(we note that animal Twitter is generally uplifting though).

For each day, we determine usage frequency for all
1-grams appearing on that day.
We also create the resultant Zipf distribution~\cite{zipf1949a},
ranking 1-grams by descending order of counts, denoting rank by $\zipfrank$.

In what follows, we first use 1-gram ranks.
As such, we do not need to be concerned with the extremely heavy tails
of frequency and Zipf distributions for Twitter,
and concomitant worries about subsampling given our
corpus's approximate 10\%-of-all-tweets character.
(i.e., we are, not unreasonably, not assured by Twitter that our
subset is exactly 10\% of all tweets).
We note that rates of 1-gram appearance for 1-grams that
are not too rare are quantities
we can measure well by simple normalization of frequencies
by the sum of all counts.
The phrase ``not too rare'' would have to be considered carefully
for studying very low fame 1-grams.
But such rates are not of importance here as, again, 
we will only work with ranks, counts, and rates of
a small set of prominent entities.

For the core of our analysis, we extract the ranks and counts for names and Twitter handles for
our six political figures (see Tab.~\ref{tab:ultrafame.entities}),
along with two 1-grams which will prove to be
of value and interest: `god' and
the Twitter handle for the K-pop band BTS, `@bts\_twt'.

The five male politicians are all dominantly referred to by
their last names, while Hillary Clinton's strongest 1-gram referent is `hillary'.
One partial reasons for this would be that
Clinton shares a last name with her husband, former
US president Bill Clinton, and the use of at least her first name has
long been a practical choice for clarity.
But referring to a person by first name versus last name 
is a not uncommon instantiation of gender bias~\cite{takiff2001a,files2017a},
and has been identified in media coverage for Clinton in the 2008 democratic primary~\cite{uscinski2011a}.
Still, for our present study, six is a small sample from
which we cannot generalize (a separate comprehensive study certainly could be a topic
of another paper);
we want to be clear that we are simply taking what the data from Twitter gives us.
We can at least note that this naming bias is not completely pervasive
with major political figures.
The 1-gram `bernie' dominates for Bernie Sanders for example.
A separate issue is McCain's first name John,
which is a poor referent.
As we will see, the movement of `john' against a background level of the name
is discernible, though this is a minor issue.
In future work, we will be able to explore 2-grams and 3-grams
but we set that analysis outside of our present scope.

To better help communicate 1-gram rank, we also determine median daily rank for
two subsets of 1-grams in 2018:
(1) Function words with median rank $\zipfrank \le 1000$;
(2) Names of countries and territories including identifiable component words (e.g., `america')
for $\zipfrank > 1000$.
For these anchor words, we do not look outside of Twitter 
because there is no one special Zipf distribution for language.
If we considered Reddit or Google Books, for example, we would use anchors
that worked within the context of those corpora.
We also acknowledge that being based on Twitter as a whole,
these ranks will tend toward a US-centric view of the world
from a particular period of history,
but we nevertheless believe they 
generally provide useful
footholds for all readers.
A few examples are:
\begin{itemize}
\item[] 
  `a' with $\zipfrank$=1,
\item[] 
  `and' with $\zipfrank$=6,
\item[] 
  `la' with $\zipfrank$=16,
\item[] 
  `there' with $\zipfrank$=162,
\item[] 
  `porque' with $\zipfrank$=323,
\item[] 
  `friend' with $\zipfrank$=539,
\item[] 
  `america' with $\zipfrank$=990,
\item[] 
  `england' with $\zipfrank$=6,718,
\item[] 
  `guatemala' with $\zipfrank$=27,775,
\item[] 
  `fiji' with $\zipfrank$=104,091,
  and
\item[]
  `niue' with $\zipfrank$=1,062,883,
  the least famous country with a four letter name~\cite{scarpino2018a-personal-communication2018}.
\end{itemize}

Finally, we operationalize the concept of ultrafame within the context of Twitter.
Broadly speaking, we will consider a 1-gram to have achieved lexical ultrafame
if it is competing with the basic function words of a language (or languages).
Upon inspection of the typical function words that tend to have the highest
daily counts, we find a remarkably stable presence
for one non-function word: `god'.
As we have done for function words and names of counties above,
we use `god' as an lexical ultrafame boundary because it is sensible within the context of Twitter.

We are not claiming that `god' is the only 1-gram that could serve this role, but it is an effective choice for offering a consistent reference point as we describe the prevalence of other 1-grams. 
Indeed, we could choose other benchmarks but our work is not meant to be an exhaustive study of establishing reference points. 

The rank for `god' hovers around 300,
showing very low volatility
(see Sec.~\ref{subsec:ultrafame.timeseries},
Fig.~\ref{fig:ultrafame.storyturbulence_twitter_all_presidents_ranks027},
Fig.~\ref{fig:ultrafame.storyturbulence_twitter_all_presidents_ranks050},
and
Tab.~\ref{tab:ultrafame.ultrafame_god} for year-scale data).
For the time period of
\firsttwitterdate--2020/12/31,
the median rank
for `god' is $r_{\textnormal{god}} = \godmedian$.
The first and third quartiles for the rank of `god'
are
\godtwofive\
and
\godsevenfive,
the 
2.5\% and 97.5\% percentiles are
\godcione\ and  \godcitwo,
and the overall high and low ranks
are
\godmin\
and
\godmax.

We will ascribe lexical ultrafame to any 1-gram with rank
$\zipfrank \le \zipfrank_{\textnormal{god}} = \godmedian.$

\section{Results}
\label{sec:ultrafame.results}

\subsection{Dynamics of lexical fame and ultrafame}
\label{subsec:ultrafame.timeseries}

\begin{figure*}[t]
  \centering	
  \includegraphics[width=\textwidth]{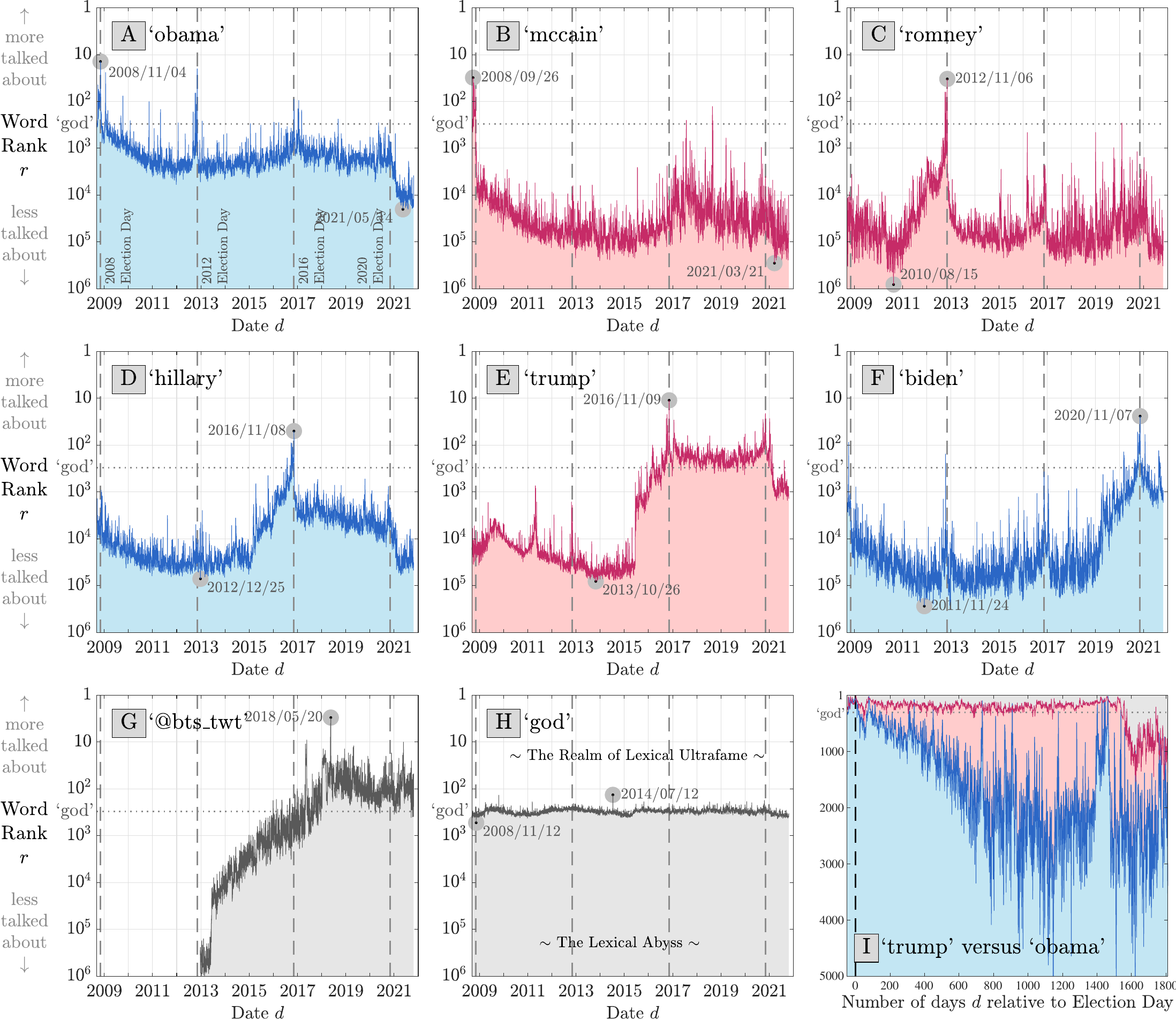}
  \caption{
    \protect
    \textbf{A--G}.
    \textbf{Temporal lexical fame on Twitter at the day scale
    for US presidents, US presidential candidates, the K-pop BTS, and the word `god'
    for the time period \firsttwitterdate\ through to \lasttwitterdate.}
    We define the lexical fame of a word as its Zipfian rank $\zipfrank$
    based on descending raw usage frequency
    (see Data and Methods, Secs.~\ref{sec:ultrafame.data} and~\ref{sec:ultrafame.methods}).
    We display lexical fame on a logarithmic scale covering six orders of magnitude.
    For the presidents and presidential candidates, we show time series of the most dominant
    word used to refer to them out of their first name, last name, and Twitter handle
    (see Sec.~\ref{subsec:ultrafame.comparisons} and
    Figs.~\ref{fig:ultrafame.storyturbulence_twitter_all_presidents_equivalences300}
    and~\ref{fig:ultrafame.storyturbulence_twitter_all_presidents_equivalences301}
    for relative usage rates).
    See Fig.~\ref{fig:ultrafame.storyturbulence_twitter_all_presidents_histograms003} for
    violin plots corresponding to time series in \textbf{A--F},
    and
    Fig.~\ref{fig:ultrafame.storyturbulence_twitter_all_presidents_ranks050}
    for the same time series presented in a wide format.
    We indicate the three US presidential elections occurring within
    the time period by vertical dashed lines, and the dates of the highest and lowest
    lexical fame on all time series.
    In all panels \textbf{A--I},
    the dotted horizontal line at a word rank of $\zipfrank = \godmedian$
    registers the global median rank
    for the word `god' through to 2021/12/31
    (panel \textbf{G}), and we consider ranks above $\godmedian$
    to be in the
    realm of lexical ultrafame.
    The time series are varied: `obama' has remained relatively famous throughout;
    `mccain' and `romney' have low, noisy fame outside of their candidacy periods;
    `hillary' has remained high post the 2016 election;
    and `trump' has achieved enduring lexical ultrafame, competing with basic function words.
    The band BTS, which most often appears through their Twitter handle, @bts\_twt, has
    followed an exponential climb into a class of lexical ultrafame unto itself, exceeding even that of `trump'.
    \textbf{I}.
    \textbf{Comparison of the lexical fame of `obama' and `trump' on Twitter
    relative to the date of their respective elections in 2008 and 2016
    as marked by the vertical dashed line at $d = 0$.}
    We also include 50 days before each election.
    Because `obama' and `trump'
    are so prominent on Twitter during these time periods,
    we are able to display word rank $\zipfrank$ on a linear scale,
    rather than the logarithmic one 
    of panels \textbf{A}--\textbf{G}.
    \todo{Fix this to go through to end of 2020 only}
    The 1-gram `trump' is remarkable for both its ultrafame level of rank
    (median of \trumpmedianrank, 2016/06/01--\lasttwitterdate)
    and consistency.
    Post inauguration, `trump' never falls below a rank of $\zipfrank = \lowtrumprank$ (which occurred on \lowtrumpdate).
    The word `obama' slowly drops in rank in the first few years of Obama's presidency
    before stabilizing, and overall shows a great deal more volatility in linear rank than `trump'.
  }
  \label{fig:ultrafame.storyturbulence_twitter_all_presidents_ranks027}
\end{figure*}

\begin{figure*}[t]
  \centering	
    \includegraphics[width=\textwidth]{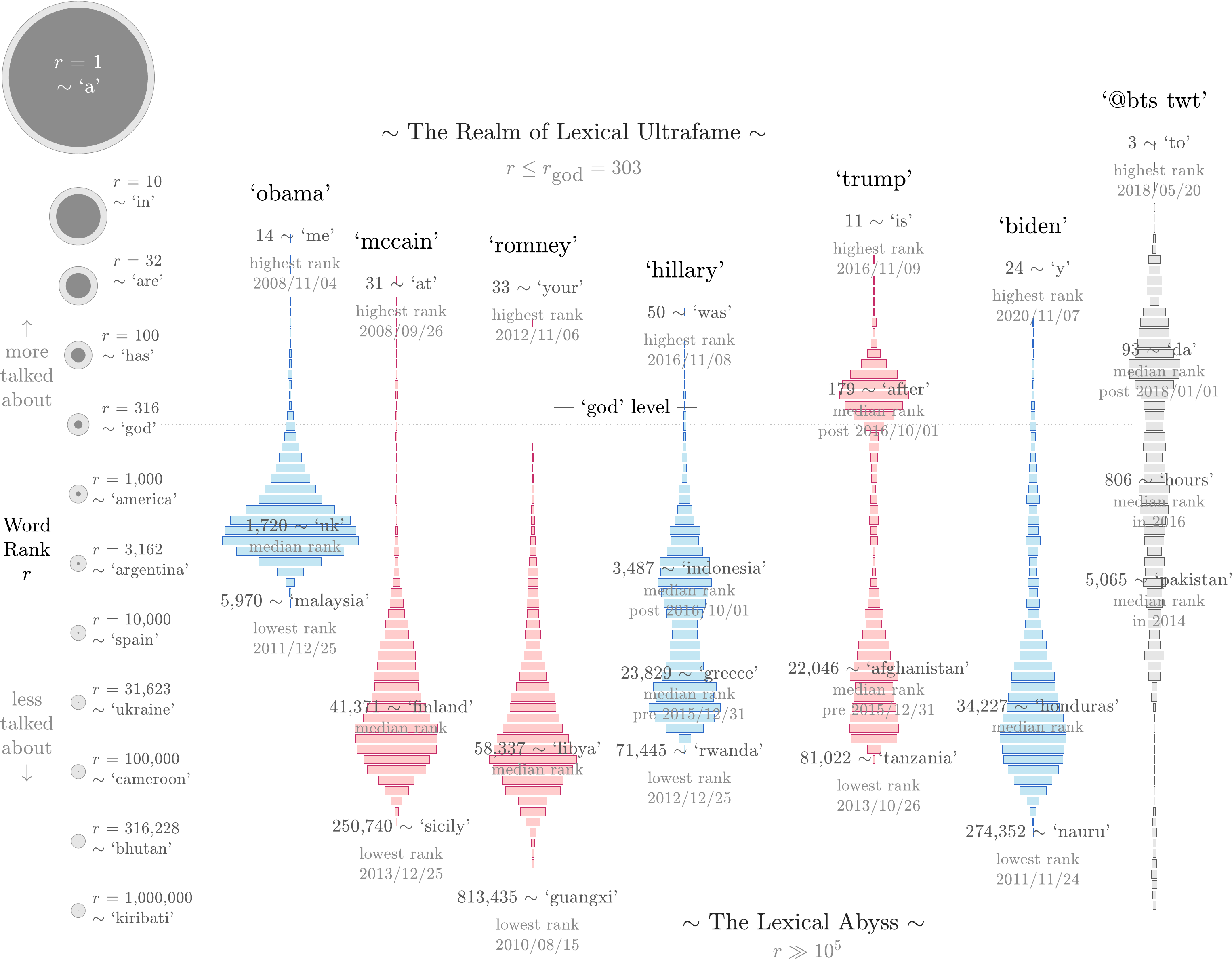}
    \caption{
      \textbf{Violin plots of lexical fame for  US presidents, US presidential candidates,
        and the K-pop BTS,}
      summarizing
      the time series
      of Fig.~\ref{fig:ultrafame.storyturbulence_twitter_all_presidents_ranks027}
      for \firsttwitterdate\ through to 2020/12/31.
      The disks on the left provide a scale for word rank at half decades,
      with the internal
      dark gray area proportional to inverse rank.
      As a guide, the example words for each disk
      are aligned with their approximate median word rank for the year 2018,
      and switch from function words (`a', `in', \ldots) to
      country or region names (`america', `argentina', \ldots).
      Consistent with Fig.~\ref{fig:ultrafame.storyturbulence_twitter_all_presidents_ranks027},
      we mark the the lexical ultrafame threshold
      with a dotted line  
      at the rank of `god' 
      (note that $\zipfrank = [10^{5/2}] = 316$
      is close to 
      $r_{\textnormal{god}} = \godmedian$).
      We indicate the highest and lowest ranks along with the dates they were attained.
      We annotate medians for the whole time period, with the exception
      of the terms `hillary' and `trump',
      for which we show medians for before 2015/12/31 and after 2016/06/01, end dates included.
      For high, low, and median ranks, we show either function or country words
      which had similar median ranks in 2018.
      For Presidents and candidates, only `trump' 
      maintains lexical ultrafame over years (median \trumpmedianrank, post 2016/10/01).
      The highest lexical fame achieved was by the Twitter handle of the band BTS,
      @bts\_twt,
      reaching a rank of 3 on 2018/05/20, matching the 2018 median rank of the word
      one of the most basic English function words: `to'.
    }
    \label{fig:ultrafame.storyturbulence_twitter_all_presidents_histograms003}
\end{figure*}

\begin{figure*}[t]
  \centering
  \includegraphics[width=\textwidth]{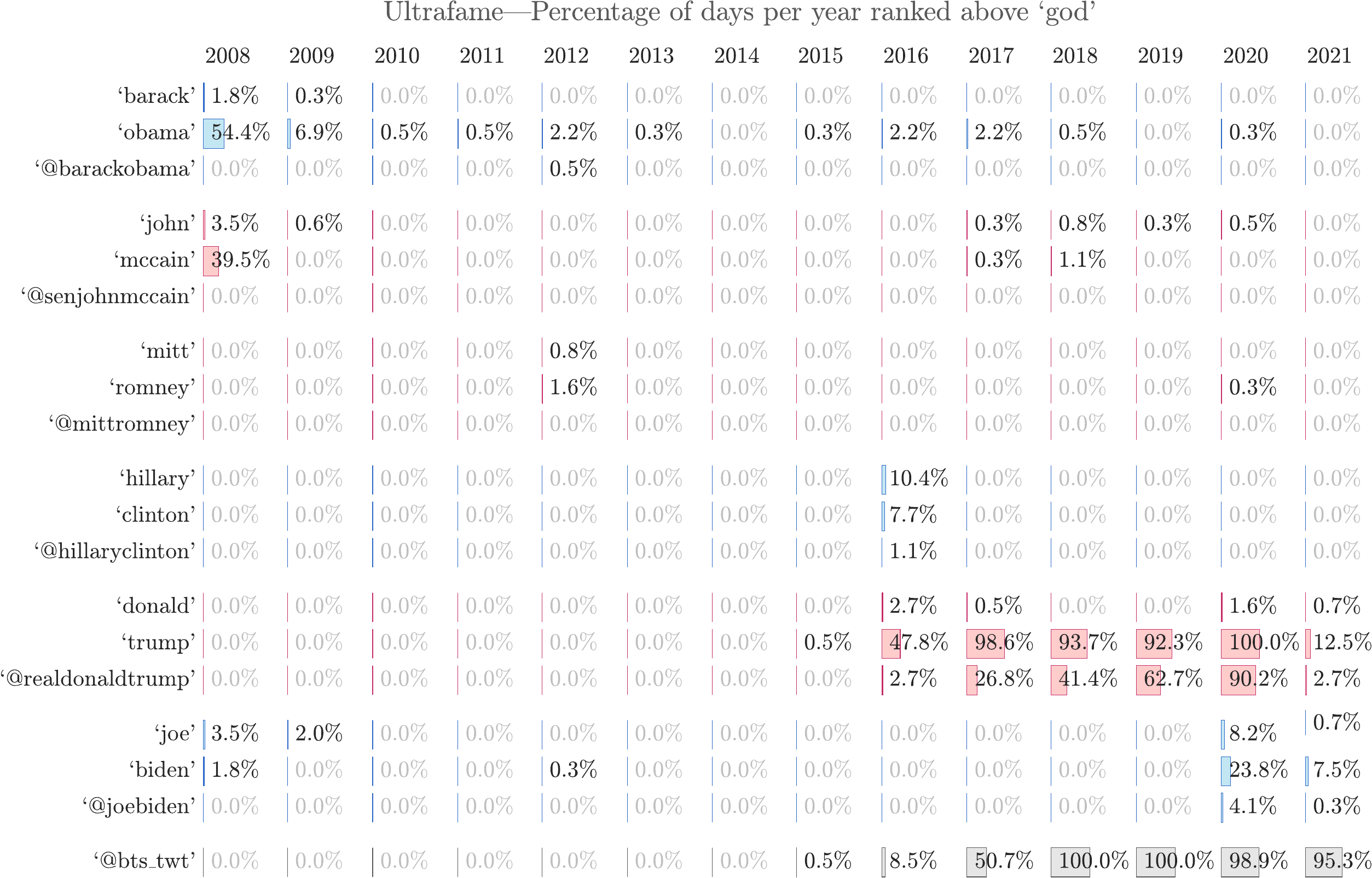}
    \caption{
      \textbf{Annual levels of ultrafame:
      Percentage of days per calendar year each 1-gram was
      ranked above (or equal to) $r_{\textnormal{god}} = \godmedian$.}
      Sustained ultrafame is rare.
      Only 1-grams associated with Trump and BTS
      have achieved enduring ultrafame across years.
      We round percentages to the nearest 0.1 percent,
      and render 0.0\% in a light gray.
      Both 2008 and \lasttwitteryear\ are for part of those years only
      (\firsttwitterdate--\lasttwitterdate, inclusive).
    }
    \label{fig:ultrafame.storyturbulence_twitter_all_presidents_ultrafame004}
\end{figure*}

We chart the 2008--2019 daily rank time series for our
six political figures, `@bts\_twt', and `god' in
Fig.~\ref{fig:ultrafame.storyturbulence_twitter_all_presidents_ranks027},
and show corresponding histograms and ultrafame rates in the companion figures,
Figs.~\ref{fig:ultrafame.storyturbulence_twitter_all_presidents_histograms003}
and
Fig.~\ref{fig:ultrafame.storyturbulence_twitter_all_presidents_ultrafame004}.
Per Tab.~\ref{tab:ultrafame.entities}, the 1-grams we track
for the political figures
are
`obama',
`mccain',
`romney',
`hillary',
and
`trump'.
We discuss these three connected figures together.

We make a number of structural elements consistent across
Figs.~\ref{fig:ultrafame.storyturbulence_twitter_all_presidents_ranks027} and
\ref{fig:ultrafame.storyturbulence_twitter_all_presidents_histograms003}.

First,
except Figs.~\ref{fig:ultrafame.storyturbulence_twitter_all_presidents_ranks027}H
and~\ref{fig:ultrafame.storyturbulence_twitter_all_presidents_ranks027}I,
we show all ranks on a logarithmic scale with limits of
$\zipfrank = 1$ and $10^6$.

Second,
in all nine plots of
Fig.~\ref{fig:ultrafame.storyturbulence_twitter_all_presidents_ranks027},
we mark the threshold of lexical ultrafame
using dotted horizontal lines at the median rank $r_{\textnormal{god}} = \godmedian$.
We visually demonstrate the stability of `god' in
Fig.~\ref{fig:ultrafame.storyturbulence_twitter_all_presidents_ranks027}G,
confirming
that `god' experiences little rank turbulence~\cite{pechenick2017a},
as reported by the statistics at the end of preceding section.
We similarly include a dotted line for the rank of `god' in
Fig.~\ref{fig:ultrafame.storyturbulence_twitter_all_presidents_histograms003}.
We more roughly locate what we call the ``lexical abyss'' in
Figs.~\ref{fig:ultrafame.storyturbulence_twitter_all_presidents_ranks027} and
\ref{fig:ultrafame.storyturbulence_twitter_all_presidents_histograms003}.
We suggest the lexical abyss
begins to appear for ranks in the hundreds of thousands,
where we have descended well below the levels populated by
commonly misspelled words to find
a wild ecology of strange lexical creatures.

Third,
we indicate US\ presidential election dates by vertical dashed lines.
In Figs.~\ref{fig:ultrafame.storyturbulence_twitter_all_presidents_ranks027}A--G,
these are for
2008/11/04,
2012/11/06,
and
2016/11/08.
In Figs.~\ref{fig:ultrafame.storyturbulence_twitter_all_presidents_ranks027}H
and~\ref{fig:ultrafame.storyturbulence_twitter_all_presidents_ranks027}I,
the `obama' and `trump' rank time series are time-shifted for direct comparison
and the day of the election is set as day number 0.

Fourth,
in Figs.~\ref{fig:ultrafame.storyturbulence_twitter_all_presidents_ranks027}A--G,
we annotate the date of the overall highest (most talked about) and lowest ranks
for the reference 1-grams.
These dates are also highlighted in
Fig.~\ref{fig:ultrafame.storyturbulence_twitter_all_presidents_histograms003},
where we provide example 1-grams typically found at those ranks.

Fifth and last,
in Appendix~\ref{sec:ultrafame.majordates},
we provide tables of extreme dates for the political figures,
`@bts\_twt', and `god'.
We list the top 10 and bottom 5 rank days for the entire time span
(Tab.~\ref{tab:ultrafame.ultrafame_overall})
as well as at the scale of each calendar year
(Tabs.~\ref{tab:ultrafame.ultrafame_obama}--\ref{tab:ultrafame.ultrafame_god}).

We discuss the time series and histograms for the six political figures and BTS
as displayed
in Figs.~\ref{fig:ultrafame.storyturbulence_twitter_all_presidents_ranks027}A--\ref{fig:ultrafame.storyturbulence_twitter_all_presidents_ranks027}F
and
Fig.~\ref{fig:ultrafame.storyturbulence_twitter_all_presidents_histograms003}
in order.
We then remark on the comparison of time series for `obama' and `trump' 
in Figs.~\ref{fig:ultrafame.storyturbulence_twitter_all_presidents_ranks027}H and
\ref{fig:ultrafame.storyturbulence_twitter_all_presidents_ranks027}I.

\bigskip

\noindent
\textbf{Lexical fame dynamics for `obama':}

\smallskip

\noindent
The lexical fame time series for `obama' can be broken
down into two main phases:
\begin{enumerate}
\item 
  Starting from a lexical ultrafame heights of Obama's 2008 campaign and election,
  a gradual decline in being talked about into 2011; and
\item 
  From the middle of 2011 to 2020/06, a largely steady state with
  an ultrafame shock for the 2012 election, a minor, years-wide cusp
  centered around the 2016 election, and a resurgence around
  the COVID-19 pandemic and the Black Lives Matter protests
  that followed the murder of George Floyd in Minnesota on 2020/05/25.
\end{enumerate}

As our historical Twitter data set begins on \firsttwitterdate,
we have on hand just short of two months of tweets leading up to
the election of Obama for his first term.
The time series for the 1-gram `obama' starts high,
achieving its highest ever rank of 
$\zipfrank$=14---a level typically held by the word `me'---attained
on the date of Obama's first election, 2008/11/04
(Fig.~\ref{fig:ultrafame.storyturbulence_twitter_all_presidents_histograms003}).

At the sub-day time scale of quarter hours,
`obama' rose to be ranked first among all words,
incredibly beating out `the' and `a'.
This peak rank for `obama' came in the 11:00 pm to 11:15 pm time frame 
on the night of the election (US Eastern Standard Time).
To reach such heights in a Zipf distribution
may seem unfathomable, and
we will offer explanations later in the paper
after we consider `@bts\_twt'.

The overall lowest rank day for `obama' was on Christmas Day in the third
year of Obama's presidency (2011/12/25)
where the 1-gram dropped to $\zipfrank=5,970$, about that of `malaysia'
(Fig.~\ref{fig:ultrafame.storyturbulence_twitter_all_presidents_histograms003}).
(Generally, we see that major holidays take precedence over politics.)
We see that after a gradual decay in rank,
`obama' resurges abruptly for the 2012 election,
and even more abruptly collapses post re-election---a spike when viewed from the level of a decade.
After level years in 2013 and 2014, `obama' slowly increases in fame again,
taking on import once again around 2016.
Post the 2016 election, `obama' has remained high in rank, showing no evident
loss of fame.

In strong contrast to the five other political figures we examine here,
lexical fame for `obama' has proved steady, durable, and relatively high on Twitter, with
a median rank of 1,720 akin to that of `uk', and a unimodal histogram
(Fig.~\ref{fig:ultrafame.storyturbulence_twitter_all_presidents_histograms003}).
But in terms of ultrafame, `obama' has been ranked above `god'
on only
\oufobama\
of all days.
Per Fig.~\ref{fig:ultrafame.storyturbulence_twitter_all_presidents_ultrafame004},
`obama' was ultrafamous on
54.4\% of the days in the last four months of 2008,
6.9\% of all days in 2009,
and then at most 2.2\% for all subsequent years.
Obama was talked about during what would be his year of re-election (2.2\%, 2012)
and then at the end of his second term (2.0\%, 2016)
and then the first year of Trump's presidency (2.2\%, 2017).
From there, the gradual decline in the rank of `obama'
(Fig.~\ref{fig:ultrafame.storyturbulence_twitter_all_presidents_ranks027}A),
has meant that in 2019 (2009/01/01 through to \lasttwitterdate),
`obama' has not once been ultrafamous.
Fig.~\ref{fig:ultrafame.storyturbulence_twitter_all_presidents_ultrafame004}
gives a first glimpse of the relative dominance of first names, last names, and
Twitter handles. We see that `barack' and `@barackobama' are well behind
the ultrafame of `obama' with only `barack' registering in 2008 and 2009.

The overall top 10 dates for `obama'
(listed in Tab.~\ref{tab:ultrafame.ultrafame_overall})
all fall on or close to election dates and inauguration dates.
High points at year scales for `obama' (Tab.~\ref{tab:ultrafame.ultrafame_obama})
are largely tied to political events even when Obama was not the central actor,
e.g., Trump's election in 2016
($\zipfrank$=90)
and inauguration in 2017
($\zipfrank$=95).

In 2018,
Obama's high ranks occurred on
September 7 ($\zipfrank$=237),
8 ($\zipfrank$=265),
and
9 ($\zipfrank$=381),
and were due to a speech he gave at 
the University of Illinois at Urbana-Champaign
where he appeared to attack President Trump
(``How hard can that be? Saying that Nazis are bad.'')~\cite{usatoday-obama-2018-09-07}.
The highest rank for `obama' in 2019 was only $\zipfrank$=685, the only year
in our data set for which Obama registered zero days of lexical ultrafame.

In 2020, `obama' has spiked on several occasions,
rising to prominence in reaction to the dominant news of the COVID-19 pandemic
and the Black Lives Matter protests.
Obama described the White House's response to the COVID-19 pandemic
as an ``absolute chaotic disaster'' in a private call~\cite{cnn-obama-2020-05-09}
on 2020/05/08, and the reporting and amplification of this
story drove Obama's name up for more than a week
with a high of $\zipfrank$=284 on 2020/05/11.
Five days after George Floyd's murder,
`obama' spiked again with $\zipfrank$=467 on 2020/07/30,
with the volume of stories around the Black Lives Matter protests
limiting the jump.
During the Democratic National Convention,
Obama's name reached highs of 
$\zipfrank$=371 on 2020/08/18
and
$\zipfrank$=350 on 2020/08/20,
with the first date being due to Michelle Obama's speech.

At the not-being-talked-about end of the spectrum,
the lowest two days for `obama' fell on New Year's Day in 2014 and 2015
($\zipfrank$=5,254 and 5,970).
Consistently across 1-grams, we see low rank days often occur on dates of major holidays
or non-political events (Tab.~\ref{tab:ultrafame.ultrafame_obama}).

\bigskip

\noindent
\textbf{Lexical fame dynamics for `mccain':}

\smallskip

\noindent
We see in Fig.~\ref{fig:ultrafame.storyturbulence_twitter_all_presidents_ranks027}B
that the time series for `mccain' encompasses four main phases:
\begin{enumerate}
\item 
  Candidate for president in 2008;
\item 
  US\ senator;
\item 
  Trump presidency;
  and
\item
  Death and legacy.
\end{enumerate}

The rank for `mccain' is highest around the 2008 election,
with a top rank of
$\zipfrank$=31 (equivalent to the usual rank of the function word `at').
Occurring on 2008/09/26, this high water mark for `mccain' 
was due to interest in the first presidential debate, held at the University of Mississippi.
In 2008, `mccain' was ultrafamous on 39.5\% of recorded days
(c.f., 54.4\% for `obama', Fig.~\ref{fig:ultrafame.storyturbulence_twitter_all_presidents_ultrafame004}).
On election day, `mccain' was still easily ultrafamous
but it would be only the 1-gram's fourth-most talked about date of the year
(2008/11/04, $\zipfrank$=54).

Across the entire time frame, McCain's 1-gram fame level is similar
to that of `finland'
($\zipfrank$=41,371)
with a low point on par with
`sicily' (2013/12/25, $\zipfrank$=250,740)
(Fig.~\ref{fig:ultrafame.storyturbulence_twitter_all_presidents_histograms003}).
Like `obama', the lexical fame of `mccain' collapsed over time renders
a unimodal histogram.
Ranking above `god' on only \oufmccain\ of all days,
outside of 2008 `mccain' was briefly ultrafamous again in only 2017 and 2018
with rates of 0.3\% and 1.1\%.

Immediately post election,
we see a sharp drop for `mccain' 
followed by a slow decay in rank over the ensuing second-phase years,
flattening out through 2013--2015.
The lowest-highest rank for `mccain' in
a calendar year
was 12,988 in 2014
(Tab.~\ref{tab:ultrafame.ultrafame_mccain}).
Throughout the Obama presidency,
McCain was often talked about when he
spoke out about decisions made by the Obama administration.
For example, the second most talked about day for `mccain' in 2016
arose on 2016/06/16 when McCain suggested that Obama's foreign policy
led to the Pulse nightclub shooting Orlando
($\zipfrank$=3,491; Tab.~\ref{tab:ultrafame.ultrafame_mccain}).

Starting in 2015, leading up to and elevating through
the 2016 election, the time series
for `mccain'
enters a third stage which is one of high fluctuations
as it begins to track McCain's increasingly
antagonistic relationship with President Trump.
The incipient event appears to have been
Trump's dismissal of McCain's record as prisoner of
war in Vietnam~\cite{nytimes2015-07-19trump-mccain}---``He's not a war hero.
He's a war hero because he was captured.
I like people who weren't captured''---which led
to the most talked about sequence of days in 2015 for
`mccain' (peak: 2015/07/18, $\zipfrank=2,557$).
(We note that some of the later mentions of `mccain' will be due to McCain's
daughter Meghan McCain, a public figure herself by this time.)

Marking the transition into a fourth phase, the highest rank for `mccain' in 2016 fell on
Election Day (2016/11/08, $\zipfrank$=2,383).
McCain had won re-election to the US senate
himself, and his withdrawn support
for Trump took on heightened salience with Trump's win.

In 2017, McCain's thumbs-down vote against in the senate to defeat a bill
to end the Affordable Care Act
generated a return to the
realm of lexical ultrafame
for the first
time since 2008 (2017/07/28, $\zipfrank$=252; Tab.~\ref{tab:ultrafame.ultrafame_mccain}).

In 2018, his death on August 25 led to a series of lexical ultrafame days
(2018/08/25, $\zipfrank$=190; 2018/08/26, $\zipfrank$=128).
The only top-10 day in 2018 that was not related to McCain's passing
came on 2018/05/11 when Kelly Sadler, a White House aide,
was reported to have said that McCain's opinion on Gina Haspel's
nomination for CIA director was irrelevant because ``he's dying anyway.''

After McCain's death, `mccain' has remained a Twitter 1-gram staple,
in part due to Trump maintaining what had emphatically
become an argument in which only one side could engage.
The high rank of $\zipfrank$=603 on 2019/03/20 was due to
Trump \cite{nytimes2019-03-20trump-mccain} who
took aim at McCain in a speech on manufacturing jobs:
``I have to be honest: I've never liked him much''~\cite{nytimes2019-03-20trump-mccain}.
In early September 2020, `mccain' reached
$\zipfrank$=955 after Jeffrey Goldberg's published article in the Atlantic
which reported that Trump had disparaged US military service members and veterans~\cite{golberg-the-atlantic2020-09-03a}.

\bigskip

\noindent
\textbf{Lexical fame dynamics for `romney':}

\smallskip

\noindent

As we did for `mccain', we can identify
four phases for the time series for `romney'
in Fig.~\ref{fig:ultrafame.storyturbulence_twitter_all_presidents_ranks027}C:
\begin{enumerate}
\item 
  Abiding in the lexical abyss: Low fame through to the end of 2010;
\item 
  A mostly uniform build to the 2012 election;
\item 
  Through to the 2016 election, a regime of fame higher and less volatile than the first phase;
  and
\item 
  A return to the lexical abyss, characteristic of the first phase.
\end{enumerate}

Romney's 1-gram has the lowest overall median rank of the six political figures,
tantamount to that of
`haiti' and again produces a unimodal histogram
($\zipfrank$=62,673, Fig.~\ref{fig:ultrafame.storyturbulence_twitter_all_presidents_histograms003}).
The low and high points for `romney' bookend the two-year-long
second phase of his fame time series.
Romney's 1-gram rose from the lexical abyss dwelling
of $\zipfrank$=813,435
(similar to `guangxi')
on 2010/08/15 
to an ultrafamous
$\zipfrank$=33
(similar to the function word `your')
on the 2012 election day (2012/11/06).

The 1-gram `romney' was ultrafamous on \oufromney\ of all days in our
study---a total of only 6 days
with
$r_{\textnormal{romney}} \le r_{\textnormal{god}} = \godmedian$.
Per Fig.~\ref{fig:ultrafame.storyturbulence_twitter_all_presidents_ultrafame004}
and Tab.~\ref{tab:ultrafame.ultrafame_overall},
all of these days occurred around the 2012 election
(1.7\% of 2012).

In the third phase of the time series,
`romney' slowly gains fame as the 2016 election approaches,
spiking on a few isolated occasions,
and then once again in 2019.
Unlike `obama' and `mccain', two of the top 10 overall days for `romney'
do not fall around election dates directly involving Romney.
In Tab.~\ref{tab:ultrafame.ultrafame_overall},
the 7th and 10th ranked days for `romney' are
2020/02/05 
and 
2019/01/02 ($\zipfrank$=457).
Further, outside of 2012, these are only two of four days on which
`romney' was ranked in the top 1000 1-grams.

Like McCain, Romney had had contentious public interactions with Trump,
and we see the cause of both of these two major spikes
was Romney speaking out against Trump.
On 2016/03/03,
Romney gave a speech at the University of Utah in which
he attacked (then potential nominee) Trump calling him a `fraud' and
asked voters to strategically vote against him~\cite{romney-speech2016-03-03}.
Almost three years later on 2019/01/01,
having been elected a senator for Utah two months prior,
Romney published a stir-causing opinion piece in the Washington Post
regarding
his negative view of Trump's character~\cite{romney2019a}.
On 2020/02/05,
Romney's lone Republican vote
in the Senate to convict Trump on one impeachment charge
again abruptly elevated Romney's name ($\zipfrank$=288).
All of these spikes evaporated, leaving no evident residual.

\bigskip

\noindent
\textbf{Lexical fame dynamics for `hillary':}

\smallskip

\noindent

The lexical fame time series for Hillary Clinton's dominant 1-gram `hillary'
in Fig.~\ref{fig:ultrafame.storyturbulence_twitter_all_presidents_ranks027}D
traverses three major phases:

\begin{enumerate}
\item 
  A gradual initial decay to a stable moderate lexical fame through the end of 2014;
\item 
  Two year build towards the 2016 election;
\item 
  A new lexical fame regime, above that of the first phase, relatively stable but gradually falling away.
\end{enumerate}

Our time frame begins after Clinton's unsuccessful campaign for
the Democratic presidential nomination
for the 2008 election, and the starting
point for `hillary' comes part way into a consequent drop in lexical fame.
As Obama's Secretary of State during his first term,
Clinton maintained a degree of public prominence that
slowed a fall towards the lexical abyss.
Until the year 2015, `hillary' was ranked in the top 1000
on a solitary day, the one on which the possibility of her
becoming Secretary of State was made public
(2008/11/14, $\zipfrank$=910, Tab.~\ref{tab:ultrafame.ultrafame_hillary}).
The nadir for the 1-gram `hillary' landed on 2012/12/25, Christmas Day
($\zipfrank$=71,445, similar to that of `rwanda';
Fig.~\ref{fig:ultrafame.storyturbulence_twitter_all_presidents_histograms003}
and
Tab.~\ref{tab:ultrafame.ultrafame_hillary}).

The second phase for `hillary' is a two-year linear ascent in
$\log_{10} \zipfrank$ starting 2015/01.
There are a few spikes during this period,
notably the day Clinton declared her candidacy
(2015/04/12, $\zipfrank$=662, Tab.~\ref{tab:ultrafame.ultrafame_hillary};
and the first Democratic debate
(2015/10/13, $\zipfrank$=446, Tab.~\ref{tab:ultrafame.ultrafame_hillary})

After a vacation lull at the end of 2015,
the first half of 2016 saw `hillary' jump and maintain a high fame level
of around a rank of $\zipfrank$=1,000.
Once Clinton became the Democratic nominee,
`hillary' once again began to climb in rank.
The first presidential debate (held at Hofstra University)
between Clinton and Trump 
led to the third highest rank overall for `hillary'
and the first time Clinton's 1-gram had achieved
lexical ultrafame
(2016/09/26, $\zipfrank$=89, Tab.~\ref{tab:ultrafame.ultrafame_overall}).

The second phase for `hillary' ends with Clinton's loss
in the 2016 election, the date of which would prove to be
the all-time top rank for `hillary'
(2016/11/08, $\zipfrank$=50, Tab.~\ref{tab:ultrafame.ultrafame_overall}).
A rank of 50---the typical return for the word `was' (Fig.~\ref{fig:ultrafame.storyturbulence_twitter_all_presidents_ultrafame004})---is lower than that achieved by both `mccain' and `romney',
and is possibly in part to `hillary' being one of several 1-gram's used to refer to Clinton.
Indeed, the day of the 2016 election would  be the high rank for `clinton' over the entire
time frame ($\zipfrank$=99).

The year 2016 is the only year in which `hillary' would achieve any
level of lexical ultrafame (10.7\% of all days in 2016, Fig.~\ref{fig:ultrafame.storyturbulence_twitter_all_presidents_ultrafame004}).
Both `clinton' and `@hillaryclinton' were also ultrafamous (7.9\% and 1.1\%).
All told, `hillary' was ultrafamous on \oufhillary\ of all days over the entire period of study,
all of them contained in 2016.

After the election, `hillary' falls abruptly, a shock transition to the third phase
of Clinton's lexical fame.
In the three following years, `hillary' trends gradually downwards,
reflected in the high ranks for 2017, 2018, 2019, and 2020:
$\zipfrank$=536 on 2017/11/03,
following book-delivered accusations by Donna Brazile that Clinton controlled the Democratic National Committee;
$\zipfrank$=713 on 2018/07/16,
apparently due to remarks by Russian President Vladimir Putin in a meeting with Trump in Helsinki, in which he stated that he wanted Trump to win;
$\zipfrank$=528 on 2019/10/19,
arising in part from the conclusion of a State Department investigation
into Clinton's email server usage,
as well as Clinton suggesting that Tulsi Gabbard was being groomed by Russia;
and
$\zipfrank$=933 on 2020/02/01,
after
Rashida Tlaib booed Clinton at a Bernie Sanders rally in Iowa.

But Clinton's lexical fame moved to a substantially higher level relative to the first phase.
The typical ranks for the first and  third phases for `hillary'
roughly differ by an order of magnitude
(see Fig.~\ref{fig:ultrafame.storyturbulence_twitter_all_presidents_histograms003}).
Before 2015/12/31,
the median rank for `hillary' was
$\zipfrank$=23,829, on par with `greece'.
Post 2016/10/01,
the median rank has elevated
to 3,139, matching the level of `argentina'.

Such a clear shift in levels of being talked about is not what we saw
for `romney' for which we have before- and after-election phases
that match in statistical character (we do not have data for `mccain' prior to the 2008 election).
Clinton has been talked about much more in a stage of her career where she
has no public position than an earlier one where she was Secretary of State for the US.

The two distinct quasi-stationary regimes for `hillary' lead
to a bimodal histogram in Fig.~\ref{fig:ultrafame.storyturbulence_twitter_all_presidents_histograms003},
in contrast to the unimodal histograms of
`obama', `mccain', and `romney'.

\bigskip

\noindent
\textbf{Lexical fame dynamics for `trump':}

\smallskip

\noindent

The top 10 ranked dates for `trump'
fall, as we might expect,
on or adjacent to
the presidential debates in 2016,
the 2016 election,
and
Trump's 2017 inauguration (Tab.~\ref{tab:ultrafame.ultrafame_overall}).
There is however much other structure to discern,
and we work through the rich details of the lexical fame time series for `trump',
which has four major phases
(Fig.~\ref{fig:ultrafame.storyturbulence_twitter_all_presidents_ranks027}E):

\begin{enumerate}
\item 
  A brief initial increase in lexical fame reaching a cusp centered around August of 2009;
\item
  A slow, birtherism-punctuated descent into the lexical abyss running into 2015;
\item
  Starting with a shock transition on 2015/06/15, an upward trajectory until the 2016 election;
\item 
  The Trump presidency, a period where
  the word `trump' has established an enduring level of
  lexical ultrafame.
\end{enumerate}

The first phase for `trump' sees a rise to a cusp point
with ranks in the 3000s.
Of the one-day spikes
(see Fig.~\ref{fig:ultrafame.storyturbulence_twitter_all_presidents_ranks027}E and Tab.~\ref{tab:ultrafame.ultrafame_overall}),
stories that failed to persist,
we see a range of causes that are political and business-related.

On 2008/10/15, `trump' reached a high for 2008 of $\zipfrank$=5,249,
following his statement in a CNN interview with Wolf Blitzer
that Nancy Pelosi should have impeached President George W. Bush
over the Iraq War~\cite{greenberg2016a}
(Trump was a registered Democrat into 2009).

The highest point for `trump' over the first few years fell on 2009/05/12,
when the 1-gram reached $\zipfrank$=1,668
after Trump
asserted that Carrie Prejean
could keep her title,
Miss California USA,
after she publically said she did not support same-sex marriage
enraging
pageant judge Perez Hilton
and, more broadly,
the internet~\cite{seelye2009a}.
Less than a month later, Trump would be involved in firing
Prejean for shirking duties set out by her contract~\cite{mckinley2009a},
making for another spike
($\zipfrank$=3,670, 2009/06/10).

The second highest rank for `trump' in 2009,
$\zipfrank$=3,227,
came on what was very much a bad news day:
On 2009/02/17, Trump Entertainment Resorts and nine
related Trump companies all filed for bankruptcy~\cite{clifford2009a}.

The cusp marking the transition from the first to the second phase for `trump'
appears to match with the 2008 Miss Universe pageant held in the Bahamas.
($\zipfrank$=3,617, 2009/08/23).
After this point, `trump' goes into a long, slow descent, interrupted by
occasional spikes and one strong resurgence which traces
Trump's major role in the birther conspiracy theory movement
which claimed that Obama was not born in the
United States~\cite{wikipedia-birtherism2019a},
and his non-unconnected  consideration
of a presidential run for the 2012 election.

In Fig.~\ref{fig:ultrafame.storyturbulence_twitter_all_presidents_ranks027}E
we see `trump' break from its downward trend in the second half of 2010,
build rapidly to a high rank of
$\zipfrank$=734 on 2011/04/27,
(the day on which Obama released his long form birth certificate),
and then drop sharply shortly thereafter.
The rank $\zipfrank$=734
would be the highest overall for `trump' until 2015.

On 2011/04/30,
Trump was extensively mocked at the White House Correspondents' Dinner
by president Obama and the host Seth Meyers.
The following day, 2011/05/01,
the rank for `trump' reached back up
to $\zipfrank$=829, the only other date `trump' would
make the top 1000 until 2015.

A few weeks later, on 2011/05/16,
Trump announced that he would not seek
the republican nomination;
`trump' jumped back up to
$\zipfrank$=1,629,
and then fell back into what would become
an ambient six-year long downward trend.
Looking back, `trump' had enjoyed half a year of being talked about
more and more, albeit with strongly disjoint story frames
of successful tycoon or blustering buffoon.

From mid 2011 to mid 2015,
`trump' inhabited the lexical abyss,
finding a `tanzania'-equivalent
low point of
$\zipfrank$=81,022 on 2013/10/26.
The lack-of-fame problems for `trump' were strongest in 2014
where the highest rank for the untalked-about `trump' was
$\zipfrank$=13,069 on 2014/09/29 (Tab.~\ref{tab:ultrafame.ultrafame_trump}).
There were a few spikes for `trump' during this time period,
two of note around the 2012 election
($\zipfrank$=1,627 on 2012/10/24
and
$\zipfrank$=1,921 on 2012/11/07).

The third phase begins with a shock transition
on 2015/06/16,
the day Trump announced his candidacy.
The 1-gram `trump' jumped from
$\zipfrank$=24,772 the day before up to 
$\zipfrank$=621.
Just five days before, `trump' was at $\zipfrank$=41,090.
From not being able to break into the top 10,000 1-grams in 2014,
`trump' now reached well inside the top 1,000 on many dates in 2015.
On 2015/12/08, `trump' was ranked 231.
Having been ranked as low as 70,230 in 2015,
`trump' would not fall below 1,297 in 2016 (Tab.~\ref{tab:ultrafame.ultrafame_trump}).

Post declaration of candidacy,
`trump' rises steadily for the next 17 months,
peaking
at $\zipfrank=12$---normally where the word `is' is---the day after the 2016 election (2016/11/09).

In 25 of 192 quarter hour intervals
on 2016/11/08 and 2016/11/09, `trump' was ranked a staggering 4th overall,
mostly in the late hours of election day and early morning hours of the following day.
The highest rank `hillary' achieved during these two days was 22nd.
This peak came in the quarter hour starting at 9 pm on the night of the election (2016/11/08),
before `hillary' began to drop down as Trump started to become perceived as the likely winner.

After a minor relative draw down post election,
`trump' surges again at Trump's inauguration
($\zipfrank$=20, 2017/01/20).
The time series for `trump' then settles into an
a scoreboard-shattering fourth phase,
a stable, low-volatility ultrafamous regime
(Fig.~\ref{fig:ultrafame.storyturbulence_twitter_all_presidents_ranks027}E).

During the fourth phase, Trump's presidency, a few dates stand out
(Tab.~\ref{tab:ultrafame.ultrafame_trump}).
These dates have largely been highly controversial, diverse in nature, and
of course, generative of enormous coverage and reaction.
We will report on distinct events that lifted `trump' to the top 10 rank
for each calendar year.

In 2017, the 8th most talked about day for `trump' and the only non-inauguration related day
in the top 10 for that year (2017/08/15, $\zipfrank$=62),
fell on the Tuesday after the Charlottesville white supremacist rally
on August 11 and 12, and the death of protester Heather Heyer on August 12.
On that Tuesday, Trump presented what would be a third statement
of his regarding the violence, a walking back of a walking back,
best captured by his assertion that there were
``very fine people on both sides''.

In 2018, `trump' peaked at $\zipfrank$=63 on 2018/07/16,
the date of the Russian-United States summit
in Helsinki when Putin expressed
his preference for Trump.

On 2018/01/12, a rank of $\zipfrank$=89 (4th highest for 2018),
followed from Trump being reported as saying that fewer immigrants should
come from ``shithole'' countries, and more from places like Norway~\cite{washingtonpost2018-01-11shithole-countries}.

On 2018/06/20
($\zipfrank$=92, 5th highest for 2018),
Trump signed an Executive Order to end
forced separation of migrant families.

The North Korea-United States summit in Singapore
was held on 2018/06/12, delivered
`trump' to a rank of
$\zipfrank$=95,
the 6th highest for 2018.

Trump's first State of the Union address on 2018/01/30
elevated `trump'
to $\zipfrank$=105, 9th highest for the year.

The highest rank day in 2019 
was 2019/12/18
($\zipfrank$=67),
and the third highest the day after ($\zipfrank$=89).

The second highest rank day in 2019
was 2019/09/25.
On 2019/09/24, Nancy Pelosi,
the Speaker of the House,
announced that impeachment proceedings would begin
against Trump.
Three relevant 1-grams that had their
highest ever ranks to date on 2019/09/25 were
`ukraine' ($\zipfrank$=264),
`transcript' ($\zipfrank$=306),
and
`whistleblower' ($\zipfrank$=700).

Earlier in the year on 2019/01/25, 
Trump signed a bill to reopen government after a prolonged
shutdown, backing down over demands to fund
the US-Mexico border wall.
On that day, `trump' reached a high that would last until the
impeachment inquiry with $\zipfrank$=112.
Three other related dates in January were also days of high ranks for `trump'.

The start of a second summit with Kim Jong Un, this time in Hanoi,
provided the another high ranked day on 2019/02/27
($\zipfrank$=114).

Trump's attacks on four congresswomen,
coupled with ``Send her back'' chants at his rallies,
provided the further highly ranked days of 2019 
(2019/07/18, $\zipfrank$=122).

One relatively low controversy date that stood out
was 2019/02/05 on which  Trump gave
his second State of the Union address,
leading to a rank of 125 for `trump'.

For the whole time span,
`trump' has been ultrafamous on \ouftrump\ of all days.
After first experiencing lexical ultrafame in 2015
(0.6\%),
the `trump' shock in 2016
lead to 49.0\% of days being ultrafamous in that year
(Fig.~\ref{fig:ultrafame.storyturbulence_twitter_all_presidents_ultrafame004}).
In 2017, 2018, and 2019,
`trump' stayed extremely high,
with ultrafame rates of
98.3\%,
93.4\%,
and
92.1\%.
Trump's low ranks during his presidency help show the persistence of fame:
$\zipfrank$=384 on 2017/12/25,
$\zipfrank$=405 on 2018/09/23,
and
$\zipfrank$=385 on 2019/09/01.
We also see the rise of @realdonaldtrump,
with ultrafame levels
in
2017, 2018, and 2019
of
26.5\%,
41.1\%,
and
60.7\%.

The year of 2020 has been unarguably tumultuous across the world.
The COVID-19 pandemic has been the globally dominant story.
In the US and with ramifications abroad,
the murder of George Floyd and the ensuing Black Lives Matter protests
have been the only other major sustained story.

In 2020, the dates surrounding the
US presidential election (2020/11/04),
the contentious first debate with Joe Biden (2020/09/29),
and Trump's contraction of COVID-19 (2020/10/02)
would prove
to be the high points for `trump' (see Tab.~\ref{tab:ultrafame.ultrafame_trump}).

Adding in the unrest leading up to and following the 2020 election,
in part instigated by Trump's rhetoric,
`trump' has been
mentioned at sustained rates higher than ever before.
After starting the year at a rank of $\zipfrank$=294
on New Year's Day,
the day-scale rank of `trump' did not fallen below 302,
meaning that `trump' sustained ultrafame for the entirety of 2020
(Tab.~\ref{tab:ultrafame.ultrafame_trump}).
Trump's handle `@realdonaldtrump' continued to grow in usage
(through retweets and mentions)
ending with an ultrafame rate of 90.2\%
(Fig.~\ref{fig:ultrafame.storyturbulence_twitter_all_presidents_ultrafame004}).

Trump's Twitter handle is the only one
for the six political figures that ever earns
sustained ultrafame.
We discuss how Twitter handles function further
below when we make sense of the fame of @bts\_twt.

We have made separate analyses
of Zipf distributions for a day of Twitter
versus the same day with retweets excluded.
While removing retweets dropped `trump' in rankings,
we observed the opposite for `@realdonaldtrump'.
Trump's Twitter handle appears then to be involved more
strongly in replies and fresh mentions than retweets.

For the start of 2021, the violent incursion into the US Capitol
by Trump supporters
has lifted rates of `trump' even higher.
The ban of Trump's account though means that his handle's
counts would be 0 from 2021/01/09 on.
Post the Capitol insurrection and ban, 
`trump' fell in rank to around 
$\zipfrank$=1000,
where it remained roughly through the end of 2021/05.
How long Trump will stay so central to the discourse on
Twitter is unpredictable.

Like `hillary',
the histogram for
`trump'
is bimodal
(Fig.~\ref{fig:ultrafame.storyturbulence_twitter_all_presidents_histograms003}).
Before 2015/12/31,
the median rank for `trump'
was $\zipfrank$=22,046, the level of `afghanistan'.

But unlike `hillary' or any of the other political figures,
the level `trump' reaches and holds in the fourth phase
of the fame time series is that of ultrafame.
From 2016/10/01 to 2020/12/31,
the median rank
for `trump'
was $\zipfrank$=179, on par with the typical rank of the word `after'.
Beyond the scales of country 1-grams,
such a rank is typical of the
function word `say'.

\bigskip

\noindent
\textbf{Lexical fame dynamics for `biden':}

\smallskip

\noindent

The 1-gram `biden' travels through three large-scale phases:
\begin{enumerate}
\item 
  An exponential fall from 2008 through 2012;
\item 
  A generally stable period of low rank from 2012 through to the end of 2018,
  with modest cusps around the 2012 and 2016
  US presidential elections
  (see Figs.~\ref{fig:ultrafame.storyturbulence_twitter_all_presidents_ranks027} and
  \ref{fig:ultrafame.storyturbulence_twitter_all_presidents_ranks050}).
\item 
  After a jump at the start of 2019, an exponential rise through to the
  2020 US presidential election, with the early suggestion of a cusp peaking
  on 2020/11/07.
\end{enumerate}

We record just a few observations regarding `biden', leaving analysis
of his social media imprint as president to future work.

The high point for `biden' comes on the Saturday (2020/11/07)
after the Tuesday 2020 US presidential election,
when the race was called in his favor by most major news outlets in.
Up until 2020, `biden' had only experienced 3 days of ultrafame.
In 2020, the median rank for `biden' was 683, akin to that of
the word `army'.
The ultrafame rate for `biden' in 2020 was 23.8\%, which is
the highest level for a full year across our data set
for all figures except Trump.

Biden's volume rises on Twitter through the Democratic primaries,
due to incidents that led to and including the
impeachment of Trump over Trump's election interference involving Ukraine,
and then again with the 2020 election itself.
The first presidential candidate debate was the overall high point until
that time for `biden', eclipsed by the post election general acknowledgment of his victory.

We note that as for the first five political figures, family members being mentioned
online add to their prominence.
For Biden, his sons Beau and Hunter and his wife Jill have
been discussed online

\bigskip

\noindent
\textbf{Lexical fame dynamics for `@bts\_twt':}

\smallskip

\noindent

We come to the extraordinary lexical fame time series 
for the seven-member K-pop band BTS, as carried by their Twitter handle, `@bts\_twt'
(Fig.~\ref{fig:ultrafame.storyturbulence_twitter_all_presidents_ranks027}F).
BTS's emergence, fame, and now central role in the K-pop industry has
been studied from a few angles by others~\cite{aisyah2017a,dal2018a,anderson2018a,wikipedia-bts-impact2019a}.
Purposefully fostered by South Korea's Ministry of Culture in the 1990s,
South Korea's development and export of cultural products led to the
``Korean Wave'' expanding across Asia in the 2000s~\cite{ryoo2009a,jin2016a},
and then a global explosion in the 2010s.
BTS's fame has translated into real money,
and they have substantially impacted the South Korean economy as well as
sales for the global music industry~\cite{wikipedia-bts-impact2019a}.

The main phases for BTS's lexical fame are:
\begin{enumerate}
\item
  Spending the first half of 2013 in the lexical abyss;
\item
  A ruthless march towards the realm of
  lexical ultrafame from mid 2013 through to the end of 2017;
\item
  Lexical ultrafame from the start of 2018 to May 2019.
\item
  A gradual drop from May 2019 while still maintaining lexical ultrafame,
  and then a resurgence in the middle of 2020.
\end{enumerate}

After breaching the top $10^{6}$ of 1-grams on 2012/12/22,
`@bts\_twt' remained in the lexical abyss and did not break
the $\zipfrank$=$10^{5}$
mark until around the time BTS
released their debut album `2 Cool 4 Skool'
and single `No More Dream'
on 2013/06/12.

Though the reception of BTS's first album and singles did not much portend for
global success---on Korean charts, `2 Cool 4 Skool' reached \#5, the lead
song would only reach \#124,
and the band's second single 
failed commercially~\cite{wikipedia-bts2019a}---the
band had entered what would become the second phase of their lexical fame.
The ascent of `@bts\_twt' to lexical ultrafame is linear in
$\log_{10} \zipfrank$ and thus exponential in $\zipfrank$
(Fig.~\ref{fig:ultrafame.storyturbulence_twitter_all_presidents_ranks027}F).

In 2014,
the median rank for `@bts\_twt'
was at the level of `pakistan'
($\zipfrank$=5,065,
Fig.~\ref{fig:ultrafame.storyturbulence_twitter_all_presidents_histograms003}).
By 2016, the median rank had climbed to 804,
on par with the word `hit'.
For 2018 and 2019,
the third phase of `@bts\_twt', 
BTS's handle
stabilized around a standard usually held
by the word `they'
($\zipfrank$=67).

The highest ranks for `@bts\_twt'
for each calendar year
even more strongly
show the explosion of BTS's lexical fame
(Tab.~\ref{tab:ultrafame.ultrafame_@bts_twt}):
$\zipfrank$=967 on 2014/12/31,
$\zipfrank$=176 on 2015/12/29,
$\zipfrank$=99 on 2016/12/29,
$\zipfrank$=9 on 2017/05/21,
to the almost incomprehensible
$\zipfrank$=3 on 2018/05/20,
and then dropping to highs of 12 and 17 in 2019 and 2020.
For comparison,
the function word `of' is on average ranked 9th,
and `to' is on average ranked 3rd,
with only `a' and `the' above at ranks 1 and 2.

If we descend below the day scale, we find that
at the level of fifteen minute time intervals on 2018/05/20,
`@bts\_twt' was in fact ranked
first overall in 17 out of 96
quarter hours.
perhaps a reflection of a truly global dedicated fan base,
perhaps a testament to the gameability of Twitter,
it remains that a non-function word being able to beat out
all other 1-grams for lexical fame on a global social media
platform with myriad competing entities
is a truly remarkable phenomenon.

For 2019,
the highest rank for `@bts\_twt'
slipped to $\zipfrank$=12 (2019/05/01).
Overall, BTS's handle has been ranked in the top 10 1-grams
on 6 days (Tab.~\ref{tab:ultrafame.ultrafame_overall}).

The calendar year lexical ultrafame rates for
`@bts\_twt'
again show their astonishing rise
(Fig.~\ref{fig:ultrafame.storyturbulence_twitter_all_presidents_ultrafame004}):
0.6\% in 2015,
8.2\% in 2016,
50.6\% in 2017,
100\% in both 2018 and 2019,
and then slightly dropping to 98.9\% in 2020.
The lowest ranks in 2018 and 2019 for `@bts\_twt'
were 267 and 257 (Tab.~\ref{tab:ultrafame.ultrafame_@bts_twt}).
Running from 2012/12/22 to \lasttwitterdate,
`@bts\_twt' was ultrafamous on \oufbtstwt\ of all days.

How can an entity compete against the most
basic function words of a language?
While Coke did once assert itself to be `it',
a marketing goal of making
the word `coke' be used as much as the word `it' 
would be (hopefully) laughed out the door.
A rank of 3 for a non-function word would
not be normal for, say, a typical book.
In Moby Dick, a deeply cetacean-rich text,
`whale' is the most frequent non-function word
and is ranked 28th.

But Twitter is a complicated melange of text.
At times, sub-populations take on the character of a chanting, echoing crowd.
In general, retweets, replies, and mentions all combine
to drive up counts of Twitter handles.
Fandoms are especially capable of
harnessing the mechanisms of social media~\cite{highfield2013a,yoon2019a}.
BTS's fan club, ARMY, is a globally formidable following,
and most tweets from the account @bts\_twt rapidly garner 
massive numbers of interactions,
far exceeding that of US political figures.

While some degree of the activity around
@bts\_twt may be algorithmic in nature---as is true for any major figures on Twitter---we
leave such quantification to other work as we are focused on the overall observables
of the system.

The high ranks of 9, 3, and 12 for
`@bts\_twt'
in 2017, 2018, and 2019
are tied to the annual Billboard Music Awards.
BTS won Top Social Artist in each of these three years,
ending Justin Bieber's
winning streak from the
award's inception in 2011 to 2016.
(Three of the five nominees in 2019
were K-pop bands.)
On the day `@bts\_twt' was ranked 3rd (2018/05/20),
the hashtag `\#ivotebtsbbmas' reached
$\zipfrank$=7 (bbmas = Billboard Music Awards),
reflecting the efforts of ARMY.
In 2019, BTS won
Billboard's music award for Top Duo/Group,
the first year in which they were nominated.

\bigskip

\noindent
\textbf{Lexical fame dynamics for `obama'} vs.\ \textbf{`trump':}

\smallskip

\noindent

In bridging to the next section on relative fame rates,
we move on from our discussion of individual fame dynamics by returning to
Fig.~\ref{fig:ultrafame.storyturbulence_twitter_all_presidents_ranks027}
to consider two side-by-side comparisons of `obama' and `trump'.

In 
Fig.~\ref{fig:ultrafame.storyturbulence_twitter_all_presidents_ranks027}H,
we overlay the lexical fame time series for `obama' and `trump'
with the first term election days for both presidents shifted to day number 0.
We show rank on a linear scale rather than logarithmic,
and include the `god' line for ultrafame once again.

Fig.~\ref{fig:ultrafame.storyturbulence_twitter_all_presidents_ranks027}H
shows that `obama' and `trump' follow time series of
divergent character.
Per our analysis of `trump', the 1-gram `trump' remains ultrafamously high through
around 1500 days,
falling below `god' on less than 10\% of days, and showing very little
volatility on a linear scale.
By contrast, `obama' falls away steadily through Obama's first term,
and shows much greater fluctuations.
We again see that `trump' is constantly a dominant feature of Twitter's story-space,
whereas `obama', while `uk'-level famous, experiences a much more variable intensity.
There are days off for `obama' but not for `trump' (and never for `@bts\_twt').

\subsection{Direct comparisons of lexical fame}
\label{subsec:ultrafame.comparisons}

\begin{figure}[tp!]
  \centering	
    \includegraphics[width=\columnwidth]{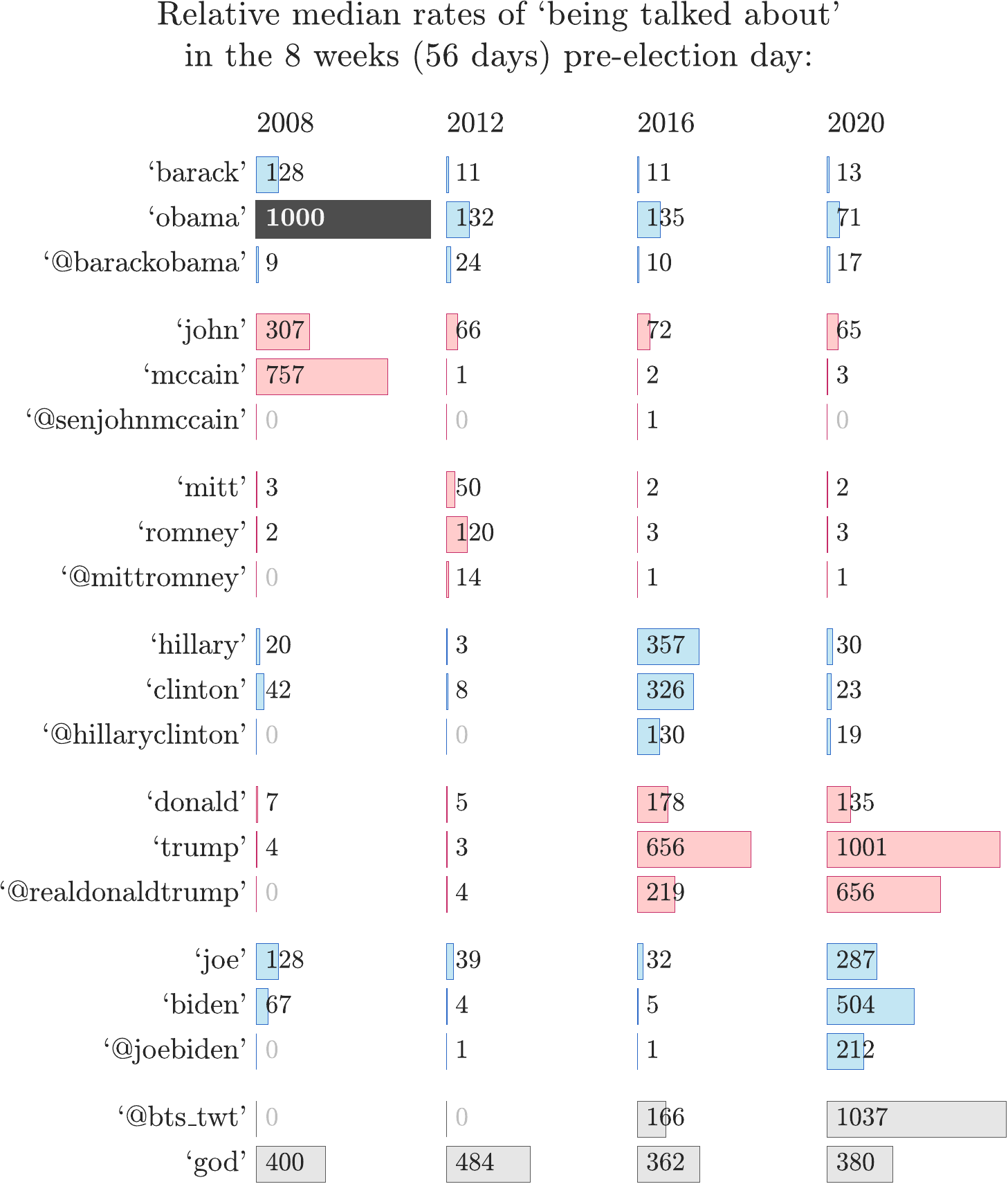}
    \caption{
      \textbf{Relative median rates of lexical fame for presidents and major candidates
        in the eight weeks leading up to the 2008, 2012, and 2016 elections}, normalized
      across terms and years using a `god' as an anchor
      (see Sec.~\ref{subsec:ultrafame.comparisons}).
      By median rate, the most talked about relatively was `obama' in 2008.
      We choose to set `obama' in 2008 to have an anchoring base rate of 1000
      (light text, dark gray bar).
      In processing a stream of tweets in a given time frame,
      for every 1000 `obama's encountered in 2008,
      all other numbers show the relative median
      rate of the number of expected mentions.
      For example, a relative count of `mccain' in 2008 shows `obama' had a roughly 4:3 advantage on Twitter (1000:757).
      The 2012 election was much less talked about with `obama' dropping to 141,
      with roughly a 9:8 advantage over his opponent `romney' (132:117).
      In 2016, `trump' outpaced `hillary' by a much stronger ratio of nearly 2:1 (656:357).
      At a relative median rate of 656, `trump' in 2016 was
      relatively less talked about than `obama' in 2008 pre-election,
      in part due to the much increased volume of Twitter.
      Over the three elections, only in 2016 did the handles of the candidates,
      `@hillaryclinton' and `@realdonaldtrump', garner substantial mentions.
      We include `@bts\_twt' and `god' for comparisons.
    }
    \label{fig:ultrafame.storyturbulence_twitter_all_presidents_equivalences300}
\end{figure}

\begin{figure*}[tp!]
  \centering	
    \includegraphics[width=\textwidth]{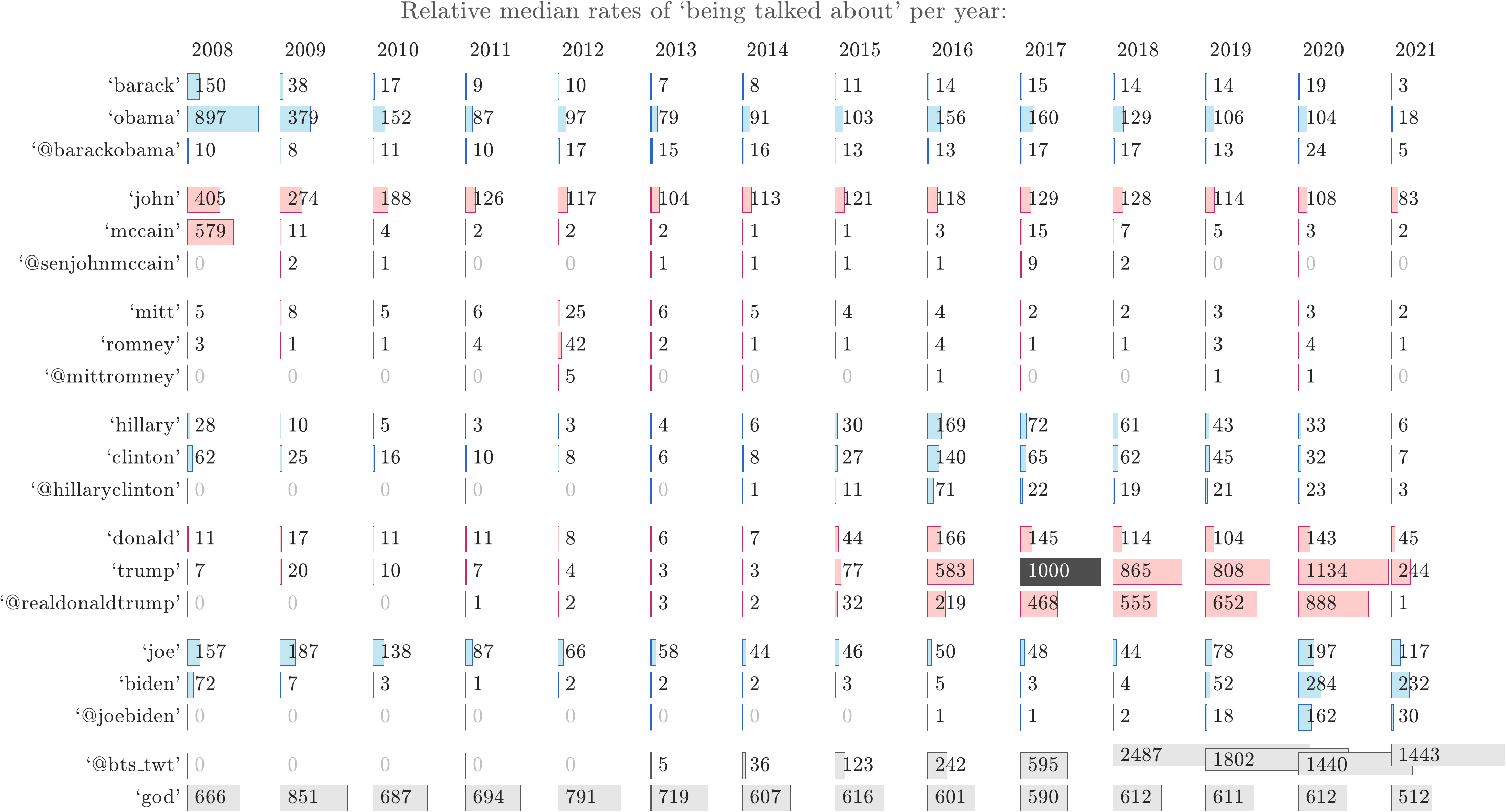}
    \caption{
      \textbf{Relative median rates of lexical fame for Presidents Obama and Trump and major candidates at the level of years.}
      Using the same approach as
      Fig.~\ref{fig:ultrafame.storyturbulence_twitter_all_presidents_equivalences300},
      we determine `trump' in 2017 to have the highest median rate and set
      that term to be the standard with a rate of 1000 (light text, dark gray bar).
      In 2018, the second year of his presidency, `trump' declined modestly to 873.
      The corresponding years for `obama' give 367 and 151, both well below `trump'
      and showing a steep first year to second year decline.
      We again include `@bts\_twt' and `god'.
      After Twitter's early growth, we see `god' stabilize from 2014 on.
      The rapid growth of `@bts\_twt' brings the handle to a median relative median rate of 2489 in 2018,
      registering an almost 3:1 ratio over `trump' in the same year (2489:865).
      We note that this relative median rate comparison figure may seem similar in appearance to
      that of Fig.~\ref{fig:ultrafame.storyturbulence_twitter_all_presidents_ultrafame004},
      but the conception and underlying calculations are different, and are worth examining separately.
      The year \lasttwitteryear\ is partial with data running through to \lasttwitterdate.
    }
    \label{fig:ultrafame.storyturbulence_twitter_all_presidents_equivalences301}
\end{figure*}

We turn now to lexical fame comparisons of 1-grams with each other and themselves,
within and across time frames.
To do so, we must move to considering normalized counts rather than ranks.
Our aim is to be able to estimate relative numbers of
mentions.
For example, in processing tweets from 2016, we want to be
able to determine how many mentions of `hillary' we should expect
for every 1000 mentions of `trump', and what would be equivalent
number of mentions of `obama' and `mccain' in 2008,
`@bts\_twt' in 2018, and so on.

We take some care to address the complications arising from
the extremely heavy-tailed Zipf distributions produced by Twitter.
With ranks, we did not have to concern ourselves with
the nature of the rare, that is,
the tails of highly skewed distributions.
(We of course derived ranks from counts which are
less informative than properly normalized relative frequencies.)
We argue that we can meaningfully interpret normalized counts as
rates rather than probabilities.

To measure rates of 1-grams, we would seem to need the
total number of 1-grams per time interval of interest,
which we will continue to take here to be the day scale.
We would appear to have a problem in that we do not know
these overall counts of all 1-grams
as we only
know that our data set comprises approximately, and not exactly,
10\% of all tweets per day.
But again this is only a problem for the tail of our distributions.
For non-rare words, we are able to accurately compute rates
by normalizing counts by the total number of 1-grams we find we have on hand per day.
More data will only modify the tails of the distributions
that we are able to deduce for our subset of all tweets
(we note that the hapax legomena for our daily Twitter Zipf distributions
are largely Twitter handles).
There are substantive ramifications here for computing
fundamental whole-distribution measures,
from simple statistics such as moments to quantities
such as the Gini coefficient and Shannon's entropy~\cite{shannon1948a}.
Here, we are able to continue on our way with our focus being on 1-grams
that are for the most part non-rare.

We generate our rate analysis in two steps.
We explain how as we present our last two main figures.
In Fig.~\ref{fig:ultrafame.storyturbulence_twitter_all_presidents_equivalences300},
for the 8 weeks leading up to the last three elections,
we show relative median rates for the first name, last name, and Twitter handle of 
all six political figures, BTS's Twitter handle,
and `god'.
In Fig.~\ref{fig:ultrafame.storyturbulence_twitter_all_presidents_equivalences301},
we show relative rates for the same 1-grams
at the scale of the calendar years 2008 through to 2019, inclusive,
and for days on which we have data.

We introduce some notation to aid our explanation.
On date $d$, we write the number of counts
of term $\tau$ as
\begin{equation}
  \termcount_{\tau,d},
  \label{eq:ultrafame.count}
\end{equation}
and the rate of term $\tau$,
the normalized count, as:
\begin{equation}
  \termratefn{\tau}{d}
  =
  \frac{
    \termcountfn{\tau}{d}
  }{
    \sum_{\tau'} \termcountfn{\tau'}{d}
  },
  \label{eq:ultrafame.rate}
\end{equation}
where the sum is over all unique terms observed on date $d$.
In general, we will use $D$ to represent a set of days with
a suitable subscript.
For example, we write $D_{y}$ for the set of days in year $y$.
So, for the year 2018,
$D_{2018} = \{2018/01/01, 2018/01/02, \ldots, 2018/12/31\}.$
For the 8 weeks leading up the US presidential election in 2016,
we would write
$D_{\textnormal{8$w$ pre-2016 election}} = \{2016/09/13, 2016/09/14, \ldots, 2016/11/07\}.$

For each 1-gram $\tau$ and each time frame $D$,
we compute the median daily rate.
(We note that the median is invariant under logarithmic transformation.)
For example, for `trump' in 2017, we would determine
\begin{equation}
  \med_{d \in D_{2017}}
  \termratefn{\textnormal{`trump'}}{d},
  \label{eq:ultrafame.ratetrump}
\end{equation}
where we use $\med$ to denote the median operator.

To now make comparisons across time periods interpretable,
we renormalize all values so that the
maximum median rate
for all political figure 1-grams is 1000, rounding to the nearest integer.
The maximum median rate is
\begin{equation}
  \termratemax
  =
  \max_{\tau,D}
  \med_{d \in D}
  \termratefn{\textnormal{$\tau$}}{d},
  \label{eq:ultrafame.maxrate}
\end{equation}
where $\tau$ and $D$ range across the political figures' 1-grams
and the time frames being compared.
So, for example, the relative median rate for `trump' in 2019 would be:
\begin{equation}
  \termrelratefn{\textnormal{`trump'}}{D_{2019}}
  =
  \frac{1000
  }{\termratemax}
  \med_{d \in D_{2019}}
  \termratefn{\textnormal{`trump'}}{d}.
  \label{eq:ultrafame.relativerate}
\end{equation}

We add that if we were interested only in comparing median rates
of 1-grams within the same time frame, we could do so without computing
rates at all.  For example, for a time frame $D$,
we could compute the median of daily ratios of counts for any two 1-grams.
It is only when we want to compare across different time ranges that we
must properly include rates.

We can now properly discuss
Figs.~\ref{fig:ultrafame.storyturbulence_twitter_all_presidents_equivalences300}
and~\ref{fig:ultrafame.storyturbulence_twitter_all_presidents_equivalences301}.
The lexical fame balance for the three pre-election periods in
Fig.~\ref{fig:ultrafame.storyturbulence_twitter_all_presidents_equivalences300}
have distinct characteristics.
For 2008, `obama' held a 1000:757
advantage over `mccain' (roughly 4:3) and both were
relatively more talked about than any other political figure later on
(first column of Fig.~\ref{fig:ultrafame.storyturbulence_twitter_all_presidents_equivalences300}).
(We note that the common name `john' was inflated by McCain's fame.)
We may speculate that `obama' led all three pre-elections in fame
in part because Twitter had a smaller user base in 2008 than in 2012 and 2016,
and so there was less competition for lexical fame across all topics.
Automatically generated content due to bots was also likely less prevalent.

For the 2012 election, `obama' again had an advantage of lexical
fame, trimmed somewhat to 132:117 over `romney'
(equivalently 1000:892, roughly 9:8,
second column of Fig.~\ref{fig:ultrafame.storyturbulence_twitter_all_presidents_equivalences300}).
A much stronger distinction is that
both `obama' and `romney' consumed far less lexical fame space,
with a relative median rate of 1000 dropping to 132 for `obama',
a factor of more than 7 lower than before the 2008 election.

Pre-election Twitter for 2016 shows `trump' outpacing `hillary' and `clinton'
by almost a 2:1 ratio at 656:357
(equivalently 1000:544,
third column of Fig.~\ref{fig:ultrafame.storyturbulence_twitter_all_presidents_equivalences300}).
Both Clinton and Trump's Twitter handles also rise in lexical fame,
in strong contrast to the handles of the other three political figures,
an effect of increased mentions and retweeting.

In short, while `trump' was relatively less talked about than `obama' in the lead up to their
respective
first elections by a 656:1000 ratio (roughly 2:3),
references to Trump dominated those of his opponent Clinton
well ahead of the smaller margins held by references to Obama over McCain and Romney.

In Fig.~\ref{fig:ultrafame.storyturbulence_twitter_all_presidents_equivalences301},
we expand the same analysis out to the full time span of our Twitter data set,
broken into calendar years.
While for the full data set, we have established much with the time series
in Fig.~\ref{fig:ultrafame.storyturbulence_twitter_all_presidents_ranks027}
and the histograms in
Fig.~\ref{fig:ultrafame.storyturbulence_twitter_all_presidents_histograms003},
the relative median rates in
Fig.~\ref{fig:ultrafame.storyturbulence_twitter_all_presidents_equivalences301}
allow us to gather further insight with hard numbers.

For political figures, the most prevalent 1-gram is now `trump' in 2017,
the first year of Trump's presidency, and we set that
median rate to a standard of 1000.

Overall, we again see the overwhelming dominance of `trump' against
the 1-grams of the other five political figures.
In the final months of 2008, `obama' performs strongly
with a relative median rate of 898,
carries a relative median rate of 379 through 2009,
but then falls and holds around a relative median rate 100 to 150 thereafter,
well below `trump'.
In 2016 as a whole, `trump' outweighed `hillary' in mentions by a ratio
of nearly 7:2 (592:170).

From 2015, Trump's Twitter handle `@realdonaldtrump' has continued to rise in lexical
fame, even while `trump' has gradually waned from the 2017 peak.
Barely apparent in 2014 with a relative median rate of 2,
`@realdonaldtrump' reaches 627 in 2019,
nearing the level of `trump' at 768.
Due to increases in mentions, retweets, and replies,
we suggest that
changes in user norms
and modifications
in Twitter's platform mechanisms
are two possible aspects of what might explain such a shift in how users reference Trump on Twitter.

Trump's abrupt rise out of the lexical abyss is again on display in
Fig.~\ref{fig:ultrafame.storyturbulence_twitter_all_presidents_equivalences301}.
From a low peak relative median rate of 20 in 2009 (relative to 1000 `trump's in 2017),
`trump' fell to 4, 3, and 3 in 2012, 2013, and 2014, barely a blip.
If the Trump's campaign goal was to Make America talk about Donald Trump Again,
then it has been a great success.

Finally, while Trump's lexical fame is clear,
BTS soundly beats all with relative median rates of `@bts\_twt'
reaching nearly 2,500 in 2018 and just below 2,000 in 2019
(in netspeak, `@bts\_twt' $>>>$ `trump').

\section{Concluding remarks}
\label{sec:concludingremarks}

We have explored in depth the daily lexical fame for six
major US political figures---Barack Obama,
John McCain,
Mitt Romney,
Hillary Clinton,
and Donald Trump---from 2008 to 2019 covering three
presidential elections.
Because of the extraordinary
being-talked-about levels that US political figures have achieved,
most especially Trump,
we have found that we needed to conceive of lexical ultrafame,
above `god' fame.
From a branding and language point of view, our findings that `trump' has been
competing with function words over the last few years should
be shocking.
As we suggested in the main text, an advertising company promising that their campaign will
elevate a brand to the level of the word `say' or `they' (par medians for `trump' and `@bts\_twt' in the last few years)---and
have days rising to compete with the word `is' and `to'---would,
we would hope, struggle to be taken seriously.

It should seem preposterous---even in the face of a global fan club,
even with the possible use
of bots and algorithmic manipulation of Twitter---that any
non-function word could be ranked third on a single day, as famous as the word `to'.
But `@bts\_twt' did just this on May 20 in 2018,
even rising to be ranked
first within quarter hour periods of the day.
The collective text of Twitter and similar kinds of social media
is distinguished from that of
other kinds of corpora
because of explicit referencing and amplification processes.
For Twitter, these processes are
automatically hyperlinked handles
and retweets.
The retweet mechanism builds in social contagion
and is adjacent to
renown (to name again)
and
reclaim (to shout again).

We have focused on one major source in Twitter for two major reasons,
and which we can now better defend.
First, Twitter provides for a measurable reflection of global events and trends,
as our ready identifications of many major events in lexical fame time series demonstrates.
Twitter is, however imperfectly, entrained with aspects of the real world.
Music and sport arguably dominate---indeed they seem to form part of a resting
state of the Twitterverse---but politics and world events are richly represented.

Second, with Twitter we have temporal resolution available in principle
at the level of a second, though here the day scale has served as our ideal time scale
for a time frame lasting over a decade.
In contrast to traditional polls, which we are in no way endeavoring
to replace but rather complement, we have a massive time series database to draw on,
which we consequently feel is deserving of a focused analysis.

In terms of future work,
we have examined in depth only lexical fame at
a daily resolution for a small set of 1-grams;
much more can be done.
Detailed investigations of thoughtfully curated sets of
competing 1-grams will always be on offer.
Other clear directions to follow
would be analogous to those taken
for search terms
by, for example, 
Google Trends
(\url{https://trends.google.com/trends/?geo=US}).
Of course, care must always be 
taken with large-scale temporal corpora
which may suffer from issues of misrepresentation (e.g., 
not including book sales or retweets),
uneven composition,
cross-contamination, 
and other problems~\cite{michel2011a,lazer2014a,pechenick2015a,alshaabi2021c}.


Considering 2-grams and 3-grams is also a natural next step,
though we caution that 2-grams and 3-grams will not 
immediately solve issues of name disambiguation.
Famous individuals are referred to in a range of ways and
comparing 1-grams with, say, 2-grams is work that must
be done with care.
Ideally, we would break language into semantically intact
phrases but we do not yet have a commonly agreed upon
approach~\cite{williams2015a}.
For the present work, we have sought to overcome
the limitations of 1-grams by considering three
essential ones for each individual:
first name, last name, Twitter handle.

We have here taken the content of all tweets in our database as
being of equal weight.
Separating out algorithmically generated tweets
would be of evident value~\cite{clark2015b,davis2016a,ferrara2016a},
as would separating retweets from ``fresh'' tweets,
and dividing tweets up by language,
and any combination of these factorings.

We emphasize that daily ranks give lower bounds on ranks for sub-day time scales.
BTS's handle `@bts\_twt' was ranked 3rd on 2018/05/20 but may have,
absurdly, held the silver or even gold medal for some period of time during that day.
More generally, a full exploration of time series at, say,
minute, 15 minutes, or hour scales for whatever topics of interest
would generate another level of fame dynamics resolution.

Finally, our work here is but one contribution to what we believe is an
emerging, post-disciplinary, data-driven science of stories.
Faithfully determining what was talked about years after the fact is an enormously challenging
enterprise in itself, and the difficulty of such
enables the intentional creation and uncontrolled emergence of false narratives.
Here, we have been able to examine a elementary part of history by following
raw Wildean fame---albeit extruded though Twitter---and thereby quantify
how much and for how long events mattered.
In the lexical fame time series of political figures,
we have seen some fundamental types of the shapes of history,
the signatures of sociotechnical time series:
stasis, noise, spikes, cusps, and shocks~\cite{dewhurst2019a}.
A data-driven categorization of the shapes
and motifs of the full ecology of rank time series
for Twitter would hold much promise for understanding
and possibly predicting
sociotechnical time series, and in the long run, stories.

\section{Data and Methods}
\label{sec:ultrafame.methods}

Our Twitter database comprises roughly 10\% of all tweets
from spans \firsttwitterdate--\lasttwitterdate.
We separate tweets into 1-grams by breaking at whitespace using a regular expression.
Certain edge cases may result in the production of 1-grams from
non-whitespace delimited sequences; these cases are relatively rare
and we did not find them to significantly affect the quality of our parsed
data.

We found we were obliged to filter out
scriptio conintua languages (languages that do not use spaces to delineate words).
we removed common characters from Japanese, Thai, Chinese, and
Korean.

After preliminary testing, we found Chinese and Japanese characters to present the
biggest challenges in terms of the high number of unique, very long
($>$ 100 characters) strings.
We accomplished the removal of these characters 
by running a regular expression to find characters in the
unicode ranges for the most commonly used Japanese and Chinese
characters.
Because character ranges for Chinese, Japanese, and Korean (CJK) are shared,
we found it necessary to remove the whole CJK range.
The regular expression we used for this step was:
\begin{verbatim}
   [\u2E80-\u2FD5\u3190-\u319f\u3400-\u4DBF
   \u4E00-\u9FCC\u4E00-\u9FFF\u3000-\u303F
   \u3040-\u309F\u30A0-\u30FF\u0E00-\u0E7F]+
\end{verbatim}

For this study, we also discarded emojis.
    
We then parse tweet bodies using a regular expression designed to capture
semantically meaningful 1-grams in a principled manner, while limiting
the artifacts of our design choices in the resulting data set. Our
regular expression for breaking on whitespace was:
\begin{verbatim}
   (https?:\/\/\w+\.\S+)|
   ([\w\@\#\'\'\&\]\*\-\/\[\=]+)
\end{verbatim}
The expression breaks down into two groups.
The first group
of the regular expression is for capturing URLs. This group captures
http and https links with arbitrary characters after the domain name
extension. The results of the URL group capture retain case
sensitivity since many links (especially those from link shortening
services) are case sensitive. The second group captures words,
hashtags, handles, and similar collections of characters. Hyphenated
words/phrases, contractions, and expressions with slashes are allowed
(thus, including many common date formats as 1-grams). We do not
impose restrictions on the number of times allowed punctuation can repeat
(e.g. ``state-of-the-art'' will be considered a 1-gram). The results of
the second group are all converted to lowercase before counting their
occurrence.

With 1-grams extracted,
we converted all Latin letters in 1-grams to lowercase.
Finally we removed the 1-grams
`rt',
`https',
`http',
`//t',
`-',
and
`t'.

We take days as based on US Eastern Standard Time.
For each day, we construct Zipf distributions by
ranking 1-grams in order of descending counts~\cite{zipf1949a}.

As we discuss in Sec.~\ref{subsec:ultrafame.analysis},
because an entity may be referred to in more than one way,
and sometimes in many ways, our simple measure of lexical fame affords a lower bound.
For example, during his two terms in office, Obama
would be indicated by `obama', `@barackobama', `\#obama', `potus', or `\#presidentobama'.
Here, we take the most common single word for each person or entity of interest.
Evidently, working with $n$-grams beyond individual terms would
allow for more complete measures of fame, and would be necessary
for names that are ambiguous referents (e.g., `bush').

For all figures, we used
MATLAB Release R2019a.

\acknowledgments
The authors appreciate comments from Ryan Gallagher, Lewis Mitchell, and Aimee Picchi.
The authors are grateful for 
support from the Massachusetts Mutual Life Insurance Company,
and the computational facilities provided by the Vermont Advanced Computing Core.
PSD and CMD were supported by NSF Grant No.\ IIS-1447634.

\clearpage

\onecolumngrid
\appendix

\renewcommand{\thefigure}{A\arabic{figure}}
\renewcommand{\thetable}{A\arabic{table}}
\setcounter{figure}{0}
\setcounter{table}{0}

\section{Time series}
\label{sec:ultrafame.timeseries}

\vfill

\mbox{}

\begin{figure*}[h!]
  \centering	
  \includegraphics[width=0.9\textwidth]{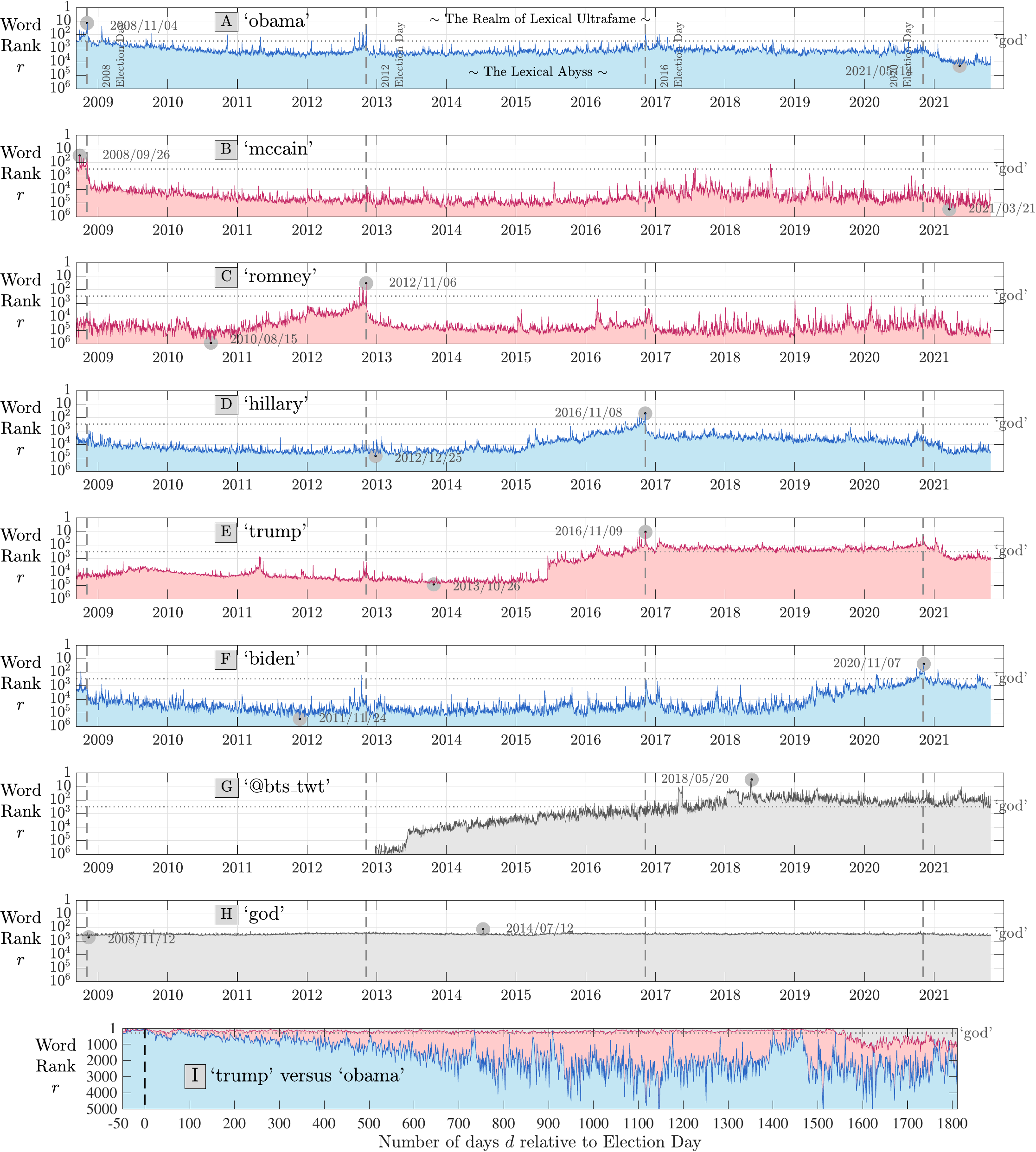}
  \caption{
    The same set of time series shown in Fig.~\ref{fig:ultrafame.storyturbulence_twitter_all_presidents_ranks027}
    rendered in wide format.
    While overall comparisons between time series may be more difficult,
    the detailed history
    presented by each time series may be more easily examined.
    Tabs.~\ref{tab:ultrafame.ultrafame_obama}--\ref{tab:ultrafame.ultrafame_god}
    list the top 10 and bottom 5 dates for each 1-gram in each year along with year-scale medians ranks.
  }
  \label{fig:ultrafame.storyturbulence_twitter_all_presidents_ranks050}
\end{figure*}

\clearpage

\section{Extreme dates}
\label{sec:ultrafame.majordates}

\begin{center}
\begin{tabular}{p{0.5\textwidth}}
    The following pages present tables showing the top 10 and bottom 5 dates
    for the ranks of the main 1-grams, first for overall period of study
    \firsttwitterdate\ through to \lasttwitterdate\
    in Tab.~\ref{tab:ultrafame.ultrafame_overall},
    and then
    at the year scale for each 1-gram
    in Tabs.~\ref{tab:ultrafame.ultrafame_obama}--\ref{tab:ultrafame.ultrafame_@bts_twt}.
  \end{tabular}  
\end{center}

\linespread{0.7}
\small

\begin{table*}[h]
  \setlength{\tabcolsep}{10pt}
  \begin{tabular}{ccc}
\textbf{`obama'} & \textbf{`mccain'} & \textbf{`romney'} \\ 
& & \\ 
1. 2008/11/04: 14 & 1. 2008/09/26: 31 & 1. 2012/11/06: 33 \\ 
2. 2008/11/05: 15 & 2. 2008/10/07: 40 & 2. 2012/10/03: 63 \\ 
3. 2012/11/06: 20 & 3. 2008/10/15: 46 & 3. 2012/10/16: 63 \\ 
4. 2009/01/20: 24 & 4. 2008/11/04: 54 & 4. 2012/10/22: 104 \\ 
5. 2008/09/26: 44 & 5. 2008/09/27: 74 & 5. 2012/11/07: 170 \\ 
6. 2008/11/03: 44 & 6. 2008/10/16: 84 & 6. 2012/11/05: 250 \\ 
7. 2012/11/07: 52 & 7. 2008/10/08: 94 & 7. 2020/02/05: 288 \\ 
8. 2008/10/07: 55 & 8. 2008/11/02: 94 & 8. 2012/10/04: 328 \\ 
9. 2008/11/01: 55 & 9. 2008/09/24: 99 & 9. 2012/10/17: 389 \\ 
10. 2008/11/02: 57 & 10. 2008/11/03: 106 & 10. 2019/01/02: 457 \\ 
$\cdots$ & $\cdots$ & $\cdots$ \\ 
4770. 2021/10/09: 18,735 & 4770. 2021/07/10: 241,159 & 4760. 2010/08/08: 449,099 \\ 
4771. 2021/07/24: 18,823 & 4771. 2021/04/18: 243,851 & 4761. 2010/11/26: 451,241 \\ 
4772. 2021/10/02: 19,236 & 4772. 2021/10/13: 244,127 & 4762. 2010/09/03: 464,323 \\ 
4773. 2021/07/31: 20,020 & 4773. 2013/12/25: 250,740 & 4763. 2010/08/11: 540,384 \\ 
4774. 2021/05/14: 20,104 & 4774. 2021/03/21: 283,921 & 4764. 2010/08/15: 813,435 \\ 
\end{tabular}
\\
  \bigskip
  \begin{tabular}{ccccc}
\textbf{`hillary'} & \textbf{`trump'} & \textbf{`biden'} & \textbf{`@bts\_twt'} \\ 
& & & \\ 
1. 2016/11/08: 50 & 1. 2016/11/09: 11 & 1. 2020/11/07: 24 & 1. 2018/05/20: 3 \\ 
2. 2016/11/09: 77 & 2. 2016/11/08: 15 & 2. 2020/11/04: 36 & 2. 2018/05/15: 4 \\ 
3. 2016/09/26: 89 & 3. 2016/11/10: 19 & 3. 2020/09/29: 52 & 3. 2018/05/14: 6 \\ 
4. 2016/10/09: 91 & 4. 2020/11/04: 21 & 4. 2020/11/06: 54 & 4. 2018/05/16: 8 \\ 
5. 2016/10/19: 110 & 5. 2016/10/09: 22 & 5. 2020/10/22: 57 & 5. 2018/05/18: 8 \\ 
6. 2016/09/27: 127 & 6. 2020/11/07: 26 & 6. 2020/09/30: 59 & 6. 2017/05/21: 9 \\ 
7. 2016/11/07: 130 & 7. 2016/09/26: 27 & 7. 2020/11/05: 59 & 7. 2021/05/23: 10 \\ 
8. 2016/10/20: 132 & 8. 2020/11/05: 27 & 8. 2021/01/20: 59 & 8. 2017/05/01: 11 \\ 
9. 2016/10/10: 138 & 9. 2021/01/06: 27 & 9. 2020/11/03: 69 & 9. 2018/05/17: 11 \\ 
10. 2016/11/10: 157 & 10. 2020/09/29: 28 & 10. 2020/11/08: 69 & 10. 2018/05/19: 11 \\ 
$\cdots$ & $\cdots$ & $\cdots$ & $\cdots$ \\ 
4770. 2011/10/09: 64,733 & 4770. 2014/08/23: 76,435 & 4770. 2012/12/25: 223,911 & 3178. 2013/04/27: 882,731 \\ 
4771. 2013/03/02: 64,804 & 4771. 2014/07/27: 77,186 & 4771. 2011/09/25: 224,846 & 3179. 2013/01/20: 882,966 \\ 
4772. 2013/12/25: 66,346 & 4772. 2013/11/17: 78,219 & 4772. 2012/04/08: 246,425 & 3180. 2013/02/10: 929,849 \\ 
4773. 2012/12/24: 67,635 & 4773. 2013/11/16: 79,322 & 4773. 2012/12/15: 263,322 & 3181. 2013/01/09: 952,557 \\ 
4774. 2012/12/25: 71,445 & 4774. 2013/10/26: 81,022 & 4774. 2011/11/24: 274,352 & 3182. 2013/02/25: 986,266 \\ 
\end{tabular}

  \caption{
    \textbf{Overall top 10 and bottom 5 rank days for the main 1-grams we track.}
    Time range covered: \firsttwitterdate\ through to \lasttwitterdate.
  }
  \label{tab:ultrafame.ultrafame_overall}
\end{table*}

\vfill

\begin{table*}
  \setlength{\tabcolsep}{4pt}
  \begin{tabular}{ccccc}
\textbf{`obama': 2008} & \textbf{`obama': 2009} & \textbf{`obama': 2010} & \textbf{`obama': 2011} & \textbf{`obama': 2012} \\ 
& & & & \\ 
\textcolor{grey}{median rank = 278} & \textcolor{grey}{median rank = 636} & \textcolor{grey}{median rank = 1,462} & \textcolor{grey}{median rank = 2,316} & \textcolor{grey}{median rank = 2,029} \\ 
\textcolor{grey}{$\sim$ `into'} & \textcolor{grey}{$\sim$ `army'} & \textcolor{grey}{$\sim$ `uk'} & \textcolor{grey}{$\sim$ `russia'} & \textcolor{grey}{$\sim$ `korea'} \\ 
& & & & \\ 
1. 2008/11/04: 14 & 1. 2009/01/20: 24 & 1. 2010/11/09: 101 & 1. 2011/05/02: 148 & 1. 2012/11/06: 20 \\ 
2. 2008/11/05: 15 & 2. 2009/01/21: 57 & 2. 2010/01/27: 256 & 2. 2011/05/01: 179 & 2. 2012/11/07: 52 \\ 
3. 2008/09/26: 44 & 3. 2009/10/09: 76 & 3. 2010/01/28: 417 & 3. 2011/01/25: 376 & 3. 2012/10/03: 66 \\
\end{tabular}

  \caption{
    \textbf{Median rank and overall top 10 and bottom 5 rank days for each calendar year for `obama'.}
    Time range covered: \firsttwitterdate\ through to \lasttwitterdate.
  }
  \label{tab:ultrafame.ultrafame_obama}
\end{table*}

\begin{table*}
  \setlength{\tabcolsep}{4pt}
  \begin{tabular}{ccccc}
\textbf{`mccain': 2008} & \textbf{`mccain': 2009} & \textbf{`mccain': 2010} & \textbf{`mccain': 2011} & \textbf{`mccain': 2012} \\ 
& & & & \\ 
\textcolor{grey}{median rank = 436} & \textcolor{grey}{median rank = 12,132} & \textcolor{grey}{median rank = 27,543} & \textcolor{grey}{median rank = 57,529} & \textcolor{grey}{median rank = 60,220} \\ 
\textcolor{grey}{$\sim$ `birthday'} & \textcolor{grey}{$\sim$ `singapore'} & \textcolor{grey}{$\sim$ `netherlands'} & \textcolor{grey}{$\sim$ `nepal'} & \textcolor{grey}{$\sim$ `libya'} \\ 
& & & & \\ 
1. 2008/09/26: 31 & 1. 2009/01/20: 2,219 & 1. 2010/02/25: 2,844 & 1. 2011/04/22: 5,475 & 1. 2012/11/14: 6,259 \\ 
2. 2008/10/07: 40 & 2. 2009/03/09: 2,308 & 2. 2010/03/26: 4,335 & 2. 2011/01/25: 7,719 & 2. 2012/11/15: 6,426 \\ 
3. 2008/10/15: 46 & 3. 2009/03/16: 2,676 & 3. 2010/01/21: 4,431 & 3. 2011/07/30: 11,911 & 3. 2012/01/04: 8,612 \\
\end{tabular}

  \caption{
    \textbf{Median rank and overall top 10 and bottom 5 rank days for each calendar year for `mccain'.}
    Time range covered: \firsttwitterdate\ through to \lasttwitterdate.
  }
  \label{tab:ultrafame.ultrafame_mccain}
\end{table*}

\begin{table*}
  \setlength{\tabcolsep}{4pt}
  \begin{tabular}{ccccc}
\textbf{`romney': 2008} & \textbf{`romney': 2009} & \textbf{`romney': 2010} & \textbf{`romney': 2011} & \textbf{`romney': 2012} \\ 
& & & & \\ 
\textcolor{grey}{median rank = 31,576} & \textcolor{grey}{median rank = 55,925} & \textcolor{grey}{median rank = 100,056} & \textcolor{grey}{median rank = 27,898} & \textcolor{grey}{median rank = 4,333} \\ 
\textcolor{grey}{$\sim$ `ukraine'} & \textcolor{grey}{$\sim$ `sudan'} & \textcolor{grey}{$\sim$ `cameroon'} & \textcolor{grey}{$\sim$ `poland'} & \textcolor{grey}{$\sim$ `colombia'} \\ 
& & & & \\ 
1. 2008/11/19: 3,575 & 1. 2009/02/28: 5,911 & 1. 2010/02/19: 8,488 & 1. 2011/10/11: 2,628 & 1. 2012/11/06: 33 \\ 
2. 2008/11/20: 8,512 & 2. 2009/03/01: 8,691 & 2. 2010/02/18: 9,627 & 2. 2011/10/18: 2,817 & 2. 2012/10/03: 63 \\ 
3. 2008/10/15: 8,557 & 3. 2009/02/06: 10,400 & 3. 2010/04/10: 10,361 & 3. 2011/12/15: 3,228 & 3. 2012/10/16: 63 \\
\end{tabular}

  \caption{
    \textbf{Median rank and overall top 10 and bottom 5 rank days for each calendar year for `romney'.}
    Time range covered: \firsttwitterdate\ through to \lasttwitterdate.
  }
  \label{tab:ultrafame.ultrafame_romney}
\end{table*}

\begin{table*}
  \setlength{\tabcolsep}{4pt}
  \begin{tabular}{ccccc}
\textbf{`hillary': 2008} & \textbf{`hillary': 2009} & \textbf{`hillary': 2010} & \textbf{`hillary': 2011} & \textbf{`hillary': 2012} \\ 
& & & & \\ 
\textcolor{grey}{median rank = 6,182} & \textcolor{grey}{median rank = 13,448} & \textcolor{grey}{median rank = 24,602} & \textcolor{grey}{median rank = 33,766} & \textcolor{grey}{median rank = 34,585} \\ 
\textcolor{grey}{$\sim$ `malaysia'} & \textcolor{grey}{$\sim$ `ecuador'} & \textcolor{grey}{$\sim$ `wales'} & \textcolor{grey}{$\sim$ `bolivia'} & \textcolor{grey}{$\sim$ `honduras'} \\ 
& & & & \\ 
1. 2008/11/14: 910 & 1. 2009/01/13: 1,669 & 1. 2010/01/21: 5,180 & 1. 2011/08/14: 3,161 & 1. 2012/12/30: 4,402 \\ 
2. 2008/11/21: 1,016 & 2. 2009/01/21: 1,925 & 2. 2010/03/03: 5,985 & 2. 2011/01/01: 7,674 & 2. 2012/10/16: 4,950 \\ 
3. 2008/12/01: 1,213 & 3. 2009/01/22: 2,049 & 3. 2010/01/16: 6,319 & 3. 2011/08/15: 8,249 & 3. 2012/12/31: 5,116 \\
\end{tabular}

  \caption{
    \textbf{Median rank and overall top 10 and bottom 5 rank days for each calendar year for `hillary'.}
    Time range covered: \firsttwitterdate\ through to \lasttwitterdate.
  }
  \label{tab:ultrafame.ultrafame_hillary}
\end{table*}

\begin{table*}
  \setlength{\tabcolsep}{4pt}
  \begin{tabular}{ccccc}
\textbf{`trump': 2008} & \textbf{`trump': 2009} & \textbf{`trump': 2010} & \textbf{`trump': 2011} & \textbf{`trump': 2012} \\ 
& & & & \\ 
\textcolor{grey}{median rank = 17,082} & \textcolor{grey}{median rank = 7,903} & \textcolor{grey}{median rank = 14,966} & \textcolor{grey}{median rank = 18,403} & \textcolor{grey}{median rank = 30,539} \\ 
\textcolor{grey}{$\sim$ `vietnam'} & \textcolor{grey}{$\sim$ `iran'} & \textcolor{grey}{$\sim$ `kenya'} & \textcolor{grey}{$\sim$ `sweden'} & \textcolor{grey}{$\sim$ `qatar'} \\ 
& & & & \\ 
1. 2008/10/15: 5,249 & 1. 2009/05/12: 1,668 & 1. 2010/01/08: 4,954 & 1. 2011/04/27: 734 & 1. 2012/10/24: 1,627 \\ 
2. 2008/09/18: 5,969 & 2. 2009/02/17: 3,227 & 2. 2010/09/09: 6,602 & 2. 2011/05/01: 829 & 2. 2012/11/07: 1,921 \\ 
3. 2008/12/01: 7,074 & 3. 2009/08/23: 3,617 & 3. 2010/01/09: 6,717 & 3. 2011/05/02: 1,097 & 3. 2012/02/02: 3,081 \\
\end{tabular}

  \caption{
    \textbf{Median rank and overall top 10 and bottom 5 rank days for each calendar year for `trump'.}
    Time range covered: \firsttwitterdate\ through to \lasttwitterdate.
  }
  \label{tab:ultrafame.ultrafame_trump}
\end{table*}

\begin{table*}
  \setlength{\tabcolsep}{4pt}
  \begin{tabular}{ccccc}
\textbf{`biden': 2008} & \textbf{`biden': 2009} & \textbf{`biden': 2010} & \textbf{`biden': 2011} & \textbf{`biden': 2012} \\ 
& & & & \\ 
\textcolor{grey}{median rank = 2,906} & \textcolor{grey}{median rank = 16,703} & \textcolor{grey}{median rank = 33,648} & \textcolor{grey}{median rank = 62,474} & \textcolor{grey}{median rank = 61,765} \\ 
\textcolor{grey}{$\sim$ `argentina'} & \textcolor{grey}{$\sim$ `vietnam'} & \textcolor{grey}{$\sim$ `bolivia'} & \textcolor{grey}{$\sim$ `haiti'} & \textcolor{grey}{$\sim$ `haiti'} \\ 
& & & & \\ 
1. 2008/10/02: 86 & 1. 2009/01/20: 775 & 1. 2010/03/23: 1,759 & 1. 2011/01/25: 4,137 & 1. 2012/10/11: 154 \\ 
2. 2008/10/03: 163 & 2. 2009/02/07: 1,772 & 2. 2010/02/14: 3,012 & 2. 2011/01/28: 6,183 & 2. 2012/10/12: 746 \\ 
3. 2008/09/26: 692 & 3. 2009/04/30: 1,805 & 3. 2010/03/24: 4,029 & 3. 2011/08/01: 9,726 & 3. 2012/09/06: 1,412 \\
\end{tabular}

  \caption{
    \textbf{Median rank and overall top 10 and bottom 5 rank days for each calendar year for `biden'.}
    Time range covered: \firsttwitterdate\ through to \lasttwitterdate.
  }
  \label{tab:ultrafame.ultrafame_biden}
\end{table*}

\clearpage

\begin{table*}
  \setlength{\tabcolsep}{4pt}
  \begin{tabular}{ccccc}
\textbf{`@bts\_twt': 2013} & \textbf{`@bts\_twt': 2014} & \textbf{`@bts\_twt': 2015} & \textbf{`@bts\_twt': 2016} & \textbf{`@bts\_twt': 2017} \\ 
& & & & \\ 
\textcolor{grey}{median rank = 22,140} & \textcolor{grey}{median rank = 5,114} & \textcolor{grey}{median rank = 1,568} & \textcolor{grey}{median rank = 806} & \textcolor{grey}{median rank = 299} \\ 
\textcolor{grey}{$\sim$ `afghanistan'} & \textcolor{grey}{$\sim$ `pakistan'} & \textcolor{grey}{$\sim$ `uk'} & \textcolor{grey}{$\sim$ `exo'} & \textcolor{grey}{$\sim$ `bts'} \\ 
& & & & \\ 
1. 2013/12/03: 4,217 & 1. 2014/12/31: 967 & 1. 2015/12/29: 176 & 1. 2016/12/29: 99 & 1. 2017/05/21: 9 \\ 
2. 2013/12/31: 4,384 & 2. 2014/12/29: 979 & 2. 2015/12/31: 260 & 2. 2016/12/26: 110 & 2. 2017/05/01: 11 \\ 
3. 2013/11/30: 4,393 & 3. 2014/12/23: 1,357 & 3. 2015/12/30: 359 & 3. 2016/11/19: 127 & 3. 2017/05/20: 13 \\
\end{tabular}

  \caption{
    \textbf{Median rank and overall top 10 and bottom 5 rank days for each calendar year for `@bts\_twt' (2013--2020).}
    Time range covered: \firsttwitterdate\ through to \lasttwitterdate.
    Note that BTS was ranked lower than $10^{6}$ on some days of 2013.
  }
  \label{tab:ultrafame.ultrafame_@bts_twt}
\end{table*}

\begin{table*}
  \setlength{\tabcolsep}{4pt}
  \begin{tabular}{ccccc}
\textbf{`god': 2008} & \textbf{`god': 2009} & \textbf{`god': 2010} & \textbf{`god': 2011} & \textbf{`god': 2012} \\ 
& & & & \\ 
\textcolor{grey}{median rank = 376} & \textcolor{grey}{median rank = 294} & \textcolor{grey}{median rank = 340} & \textcolor{grey}{median rank = 329} & \textcolor{grey}{median rank = 273} \\ 
\textcolor{grey}{$\sim$ `vote'} & \textcolor{grey}{$\sim$ `bts'} & \textcolor{grey}{$\sim$ `vote'} & \textcolor{grey}{$\sim$ `bts'} & \textcolor{grey}{$\sim$ `were'} \\ 
& & & & \\ 
1. 2008/11/05: 239 & 1. 2009/10/20: 136 & 1. 2010/04/26: 229 & 1. 2011/03/11: 188 & 1. 2012/11/22: 205 \\ 
2. 2008/12/18: 256 & 2. 2009/03/31: 194 & 2. 2010/09/11: 229 & 2. 2011/09/11: 215 & 2. 2012/12/24: 209 \\ 
3. 2008/12/25: 269 & 3. 2009/07/04: 202 & 3. 2010/04/25: 236 & 3. 2011/12/25: 226 & 3. 2012/02/18: 210 \\
\end{tabular}

  \caption{
    Time range covered: 2013/01/01 through to \lasttwitterdate.
    \textbf{Median rank and overall top 10 and bottom 5 rank days for each calendar year for `god'.}
  }
  \label{tab:ultrafame.ultrafame_god}
\end{table*}

\end{document}